\documentclass[twocolumn,a4paper,reprint,amssymb,aps,prd,groupedaddress,superscriptaddress,nofootinbib]{revtex4-2}
\pdfoutput=1

\usepackage{graphicx}
\usepackage{dcolumn}
\usepackage[colorlinks = true,
            linkcolor = blue,
            urlcolor  = blue,
            citecolor = blue,
            anchorcolor = blue]{hyperref}

\usepackage{amsmath}
\usepackage[capitalize]{cleveref}
\usepackage{amsfonts}
\usepackage{resizegather}
\usepackage{color}
\usepackage{subfigure}
\usepackage{ulem}
\usepackage{blindtext}
\usepackage{url}
\usepackage[dvipsnames,table]{xcolor}
\usepackage{lipsum}
\usepackage{graphicx}
   \usepackage{float}
\usepackage{array,mathtools,amssymb,booktabs}
\newcolumntype{C}{>{$}c<{$}}

\definecolor{violet}{rgb}{0.56, 0.0, 1.0}

\newcommand{\wde}[1]{$w_{\rm DE}(z)$#1}

\begin{document}

\title{
Model-independent reconstruction of the Interacting Dark Energy Kernel:\\
Binned and Gaussian process}

\author{Luis A. Escamilla}
\email{luis.escamilla@icf.unam.mx}
\affiliation{Instituto de Ciencias F\'isicas, Universidad Nacional Aut\'onoma de M\'exico, 
Cuernavaca, Morelos, 62210, M\'exico}

\author{\"{O}zg\"{u}r Akarsu}
\email{akarsuo@itu.edu.tr}
\affiliation{Department of Physics, Istanbul Technical University, Maslak 34469 Istanbul, Turkey}

\author{Eleonora Di Valentino}
\email{e.divalentino@sheffield.ac.uk}
\affiliation{School of Mathematics and Statistics, University of Sheffield, Hounsfield Road, 
Sheffield S3 7RH, United Kingdom}

\author{J. Alberto Vazquez}
\email{javazquez@icf.unam.mx}
\affiliation{Instituto de Ciencias F\'isicas, Universidad Nacional Aut\'onoma de M\'exico, Cuernavaca, 
Morelos, 62210, M\'exico}

\begin{abstract}

The cosmological dark sector remains an enigma, offering numerous possibilities for exploration. One particularly intriguing option is the (non-minimal) interaction scenario between dark matter and dark energy. In this paper, to investigate this scenario, we have implemented Binned and Gaussian model-independent reconstructions for the interaction kernel alongside the equation of state; while using data from BAOs, Pantheon+ and Cosmic Chronometers. In addition to the reconstruction process, we conducted a model selection to analyze how our methodology performed against the standard $\Lambda$CDM model. The results revealed a slight indication, of at least 1$\sigma$ confidence level, for some oscillatory dynamics in the interaction kernel and, as a by-product, also in the DE and DM. A consequence of this outcome is the possibility of a sign change in the direction of the energy transfer between DE and DM and a possible transition from a negative DE energy density in early-times to a positive one at late-times. While our reconstructions provided a better fit to the data compared to the standard model, the Bayesian  Evidence showed an intrinsic penalization due to the extra degrees of freedom.  Nevertheless these reconstructions could be used as a basis for other physical models with lower complexity but similar behavior.
\end{abstract}

\maketitle

\section{INTRODUCTION}\label{section:intro}

Over the course of more than two decades there has been an avalanche of theoretical studies and 
data analysis to understand the fundamental nature of the dark energy (DE), nevertheless so far it still 
remains as an open question. This mysterious component of the universe was initially introduced as a 
possibility to explain the current accelerated expansion of the Universe, discovered through the Type 
Ia Supernovae (SN) observations~\cite{SupernovaSearchTeam:1998fmf, SupernovaCosmologyProject:1998vns, Amanullah:2010vv,
SupernovaCosmologyProject:1997zqe}, and then confirmed by different measurements, like the Cosmic 
Microwave Background (CMB) anisotropies~\cite{Planck:2015mrs, WMAP:2012fli, WMAP:2010qai, Planck:2018vyg} 
or the Baryon Acoustic Oscillations (BAO)~\cite{2013PhR...530...87W, Li:2012dt, Reid:2009xm, SDSS:2009ocz, 
SDSS:2008tqn}. 
In its simplest form, the DE is assumed to be a cosmological constant ($\Lambda$) incorporated into the 
Einstein field equations (EFEs). It is equivalent to the usual vacuum energy density of the quantum field theory, 
described as a perfect fluid with a barotropic equation of state parameter (EoS) $w_{\mathrm{DE}} = p/\rho$, 
with $w_{\Lambda} = -1$. The cosmological constant, along with the cold dark matter (CDM), a key component 
for structure formation in the Universe, plays a crucial role in setting up the standard cosmological model, 
best known as Lambda Cold Dark Matter ($\Lambda$CDM) model. Even though this model is able to explain, with 
great accuracy, most of the contemporary observations, it exhibits some issues on theoretical grounds, 
like the Cosmological Constant problem~\cite{Weinberg:1988cp, 
Sahni:2002kh, Wang:1999fa, Sahni:2002kh} and the Coincidence problem~\cite{Sahni:2004ai, Alam:2004jy, 
Bull:2015stt}; and also faces some difficulties on the observational side, viz., the $H_0$
tension~\cite{Verde:2019ivm,Riess:2019qba,DiValentino:2020zio, DiValentino:2021izs,  
Kamionkowski:2022pkx, Riess:2021jrx,DiValentino:2022fjm}, and the $\sigma_8-S_8$ tension~\cite{DiValentino:2020vvd, 
Douspis:2018xlj} (see also~\cite{Abdalla:2022yfr} for a review of the current tensions and anomalies 
in cosmology).
In order to explain the small scale structure formation, or at least to ameliorate the problems associated 
with the CDM model, several alternatives have been introduced. One viable option 
is to replace the standard CDM with Scalar Field dark matter components~\cite{ideas1,ideas2,ideas4, 
Matos09, Tellez-Tovar:2021mge, Navarro-Boullosa:2023bya} or by introducing the Self Interacting dark matter~\cite{sidm, 
Gonzalez:2008wa, Padilla:2019fju}, or to support the warm dark matter scenario~\cite{Pan:2022qrr,Naidoo:2022rda}. 
Regarding the current accelerated expansion of the Universe, a natural extension to the constant EoS is 
introducing phenomenological dynamics to model the general behavior of the DE, whose main 
methodology relies on giving a functional form with a dependence on redshift/scale factor, i.e., $w_{\mathrm{DE}}=w(z)$, 
see, e.g.,~\cite{AlbertoVazquez:2012ofj, DiValentino:2017zyq, Hee:2016nho, Zhao:2017cud, Wang:2018fng}.
For instance, parametric forms of the EoS parameter $w_{\mathrm{DE}}$ have been studied extensively throughout 
several papers, as they served as guidelines to uncover some underlying issues from the theory, and are commonly 
referred to as Dynamical Dark Energy (DDE) parameterizations. One of the simplest descriptions of $w_{\mathrm{DE}}$ 
is given by a Taylor series in terms of the scale factor $a$, and a set of free parameters, i.e., $w_0$ and $w_a$, whose particular 
cases are the $w$CDM model, $w_{\mathrm{DE}} = w_0$, and the Chevallier-Polarski-Linder (CPL) parametrization, 
$w_{\mathrm{DE}} = w_0 + w_a\left(1-a\right)$~\cite{Chevallier, Linder:2002et,CPL1, CPL2, 2012comprehensive}, or 
it could be carried out in terms of redshift $z$~\cite{Zhang:2011xx}, or cosmic time $w_{\mathrm{DE}} = w_0 + 
w_a(1-t)$~\cite{Akarsu:2015yea}, or in general by using another basis of series expansion, e.g., a 
Fourier-base~\cite{Tamayo:2019gqj}. There are also more complex parameterizations that may include combinations 
of power laws, exponentials, logarithms and trigonometrics components~\cite{Linder:2005dw, Kurek:2007bu, 
Sahni:2006pa, Arciniega:2021ffa, Jaime:2018ftn,Yang:2021flj}.
Studies of the equation of state have been very useful to describe the DE features, nevertheless, several works 
have extended the search by looking for deviations from the constant energy density $\rho_{\mathrm{DE}}$. 
Examples of these investigations include the Graduated dark energy (gDE)~\cite{Akarsu:2019hmw, Acquaviva:2021jov, Akarsu:2021fol, Akarsu:2022typ, Akarsu:2023mfb}, 
the Phantom crossing~\cite{DiValentino:2020naf}, Omnipotent dark energy~\cite{Adil:2023exv} or the Early dark energy~\cite{Doran:2006kp}, just 
to mention a few.
Also, some of the possibilities that replace the cosmological constant include canonical scalar fields 
like quintessence, phantom or a combination of multi-fields named quintom models~\cite{Vazquez:2023kyx, Caldwell:1999ew, 
Caldwell:2005tm, Vazquez:2020ani, Akarsu:2020pka, quintessence, modelsDE, quintom}; or it could even be 
modifications that go beyond the General Relativity such as $f(R)$ theories~\cite{Felice100,Nojiri:2010wj} 
or braneworld models~\cite{Ida:1999ui, Brax:2004xh, modifiedgravity, Miguel18}.

On the other hand, there exists a particular type of models where the non-minimal interaction (hereinafter 
we use the word interaction to mean non-minimal interaction) between DM and DE may be able to solve or at 
least alleviate these issues with relative ease~\cite{Kumar:2017dnp, Kumar:2019wfs,  
DiValentino:2017iww, Pourtsidou:2016ico, An:2017crg, Cai:2004dk, delCampo:2008jx, Yang:2018uae, Yang:2018ubt, 
Pan:2019gop, DiValentino:2019ffd,DiValentino:2019jae,Gomez-Valent:2020mqn, Lucca:2020zjb, Bernui:2023byc, 
Nunes:2022bhn, Nunes:2021zzi}. Recent analyses have focused on the resemblance between DM-DE interacting 
models with modified gravity theories~\cite{Wang:2016lxa, Kofinas:2016fcp, Johnson:2020gzn, Johnson:2021wou}.
Also, a viable alternative is to assume a  interaction between these two
components~\cite{Brax:2006kg,Bolotin:2013jpa}.   

Even though the interacting models have been extensively analyzed, their interaction kernel still remains a mystery. 
This is why a significant amount of research works has been dedicated to introducing new models in a phenomenological way, such as the parameterizations. Models with interacting dark sectors, also named Interacting Dark Energy (IDE), are no strangers to 
parameterizations, since the interplay is generally proposed in a particular demeanor motivated by certain characteristics. For example, a popular assumption, inspired by several behaviors in particle physics~\cite{Pan:2020zza}, is to express the interaction kernel, $Q$, in terms of the energy densities ($\rho_{\rm DM}$ 
and/or $\rho_{\rm DE}$) and time (through the Hubble parameter $H^{-1}(z)$). Nevertheless, these are only few 
assumptions and, since the nature of the interaction is still obscure, they can come up in several different 
functional forms and combinations, see for example~\cite{Kumar:2019wfs,  DiValentino:2017iww, Wang:2016lxa,
Yang:2018uae,Pan:2019gop,Yang:2022csz}. 
Generally, for these type of models, it is found that the structure formation remains unaltered and late-time 
acceleration is also in accordance with the standard model (see~\cite{Wang:2016lxa, Bolotin:2013jpa} for 
a comprehensive review of interacting models and their behavior). 

However, as useful as parameterizations may be (not only for DDE or IDE but in general) they 
posses certain limitations. One of this is that a functional form is assumed a priori which could bias the results. 
For example selecting a phantom or quintessence like component remarks a clear difference in the DE 
behavior when choosing a particular parameterization. {Another limitation was demonstrated in~\cite{Colgain:2021pmf}; expansions in small parameters are more influenced by higher redshift data, whereas data from lower redshifts carry less weight in the analysis.} A possible way to avoid these issues it to perform reconstructions by extracting information directly 
from the data,  using model-independent techniques or non-parametric ones, such as Artificial Neural 
Networks~\cite{Wang:2019vxv, Gomez-Vargas:2021zyl, Benisty:2022psx}, Gaussian Process~\cite{Keeley:2020aym, Seikel2012,
Holsclaw2010, Liao:2019qoc, Liao:2020zko, Benisty:2020kdt, Benisty:2022psx, Calderon:2022cfj, Calderon:2023msm} or, recently, we can see applications of binning, linear interpolations and the incorporation 
of a correlation function in~\cite{Escamilla:2021uoj}. 
The Gaussian process (GP), specifically for the IDE models, has become a regular choice for a non-parametric
approach~\cite{Aljaf:2020eqh, Mukherjee:2021ggf, Yang:2015tzc, Bonilla:2021dql, Bonilla:2020wbn,
vonMarttens:2020apn}. This methodology has found a possibility of an interaction and given some insights 
into possible preferred behaviors and characteristics, such as a crossing of the non-interacting line. 
In spite of this, the GP approach cannot be used for model comparison in concordance with the $\Lambda$CDM 
model given its non-parametric nature.

Despite the extensive study of both the interaction models and the model-independent approaches in cosmology,  
they have been rarely used in tandem, at least to our knowledge; for example Cai et. al.~\cite{Cai:2009ht} 
and Salvatelli et. al.~\cite{Salvatelli:2014zta} used redshift bins, and for Solano et. al.~\cite{Solano:2012zw}
the main focus are the Chebyshev polynomials. 
They found a possible crossing in the non interaction line, and in~\cite{Cai:2009ht} obtained an oscillatory
behavior through the interaction, although the data in this work was limited to cover a narrow range of 
redshift (around $z<1.8$). This finding inspired the study of possible sign-switching interactions, instead 
of the classical monotonically decreasing or increasing parameterizations. We have for example: 
in~\cite{Li:2011ga} the parameterization $Q(a) = 3b(a)H_0\rho_0$ was first proposed, with $b(a) = b_0a 
+b_e(1-a)$ being the sign-switching part; in~\cite{Sun:2010vz} the model $Q = 3H\sigma(\rho_{\rm DE} - 
\alpha\rho_{\rm DM})$ ($\alpha$ being a positive constant of order unity) also presents a switching 
interaction; in~\cite{Zadeh:2017zke} the named Ghost dark energy is used in tandem with an interaction kernel $Q = 
3\beta H q (\rho_{\rm DE}+\rho_{\rm DM})$  where its sign is able to change since it is a function 
of the deceleration parameter $q$; and in~\cite{Guo:2017deu} a bunch of variations of $Q(a) = 3b(a)H(a)\rho_i$ 
are studied. The general consensus reached by the majority of these models is that, 
if a transition were to happen, it should be around the time when the accelerated expansion of the 
Universe began ($z\sim0.5$). 
Therefore, in this work we will use some model-independent approaches to reconstruct the 
interaction kernel between DE and DM directly from the data. The methods used (as will be explained in the 
next sections) are the binning scheme along with the  Gaussian Process as an interpolation approach. Moreover, 
as additional cases, together with the DM-DE interaction, we will replace the cosmological constant with a 
constant EoS free to vary.

The paper is organized as follows: in~\cref{section:ide_model} we provide a brief review of the 
underlying theoretical reason that led us to consider the possibility of non-minimal interaction within the dark 
sector, followed by~\cref{section:bin_gp_interpolation} where we describe the reconstruction methodologies.
In~\cref{section:data} the datasets and some specifications about the parameter estimation and model 
selection are made clear. In~\cref{section:results} we present 
the main results, and finally in~\cref{section:conclusions} we give our conclusions. 

\section{INTERACTING DM-DE MODEL}\label{section:ide_model}

In the general theory of relativity (GR), the Einstein field equations can be written as $G_{\mu\nu}=
\kappa T_{\mu\nu}^{\rm tot}$, where $\kappa=8 \pi G$ ($G$ is Newton's constant), $G_{\mu\nu}$ is the Einstein 
tensor, and $T_{\mu\nu}^{\rm tot}$ is the total energy-momentum tensor (EMT), viz., the sum of the EMTs 
of sources such as radiation (photons and neutrinos), baryons, CDM/DM, and DE, which constitute the physical 
content of the universe. It is an important feature of the EFEs that the twice contracted Bianchi 
identity, $\nabla^{\mu} G_{\mu\nu}=0$, implies the conservation of the \textit{total} EMT, i.e., 
$\nabla^{\mu} T_{\mu\nu}^{\rm tot}=0$. Accordingly, in a relativistic cosmological model assuming the 
spatially flat Robertson-Walker spacetime, in the presence of sources in the standard model of particle 
physics---i.e., baryons ($w_{\rm b}=0$), radiation (photons and neutrinos) ($w_{\rm r}=\frac{1}{3}$)---and 
sources of unknown nature---CDM\footnote{Since in this work we consider a non-minimal interaction between DE 
and cold dark matter (CDM) with $w_{\rm CDM}=0$, it would be more appropriate to call it only 
\textit{dark matter} (DM). Actually, in the traditional definition the CDM is supposed to interact 
only gravitationally.} ($w_{\rm CDM}=w_{\rm DM}=0$) and DE ($w_{\rm DE}$ is left unspecified)---the 
EFEs lead to the following Friedmann and continuity equations, respectively:
\begin{align}
    3H^2=\kappa(\rho_{\rm r}+\rho_{\rm b}+\rho_{\rm DM}+\rho_{\rm DE}),\\
    \label{eqn:cont}
\begin{aligned}
    \dot{\rho}_{\rm r}+4H\rho_{\rm r}+\dot{\rho}_{\rm b}&+3H\rho_{\rm b}+\dot{\rho}_{\rm DM} 
        +3H\rho_{\rm DM}\\
    &+\dot{\rho}_{\rm DE}+3H\rho_{\rm DE}(1+w_{\rm DE})=0,
\end{aligned}
\end{align}
where $H$ is the Hubble parameter, and a dot denotes derivative with respect to cosmic time. It is reasonable to 
assume that the sources such as baryons and radiation, whose physics are well known within the standard model 
of particle physics, are individually conserved, i.e., $\nabla^{\mu}T_{\mu\nu}^{\rm r}=0$ and 
$\nabla^{\mu}T_{\mu\nu}^{\rm b}=0$ (viz., $\dot{\rho}_{\rm r}+4H\rho_{\rm r}=0$ and $\dot{\rho}_{\rm b}
+3H\rho_{\rm b}=0$). This in turn implies, via continuity equation~\cref{eqn:cont}, conservation within the 
dark sector (DM+DE) itself:
\begin{equation}
\label{eqn:totconser}
    \dot{\rho}_{\rm DM} +3H\rho_{\rm DM}+\dot{\rho}_{\rm DE}+3H\rho_{\rm DE}(1+w_{\rm DE})=0.
\end{equation}
At this stage, in the cosmology literature so far, the very strong assumption that DM and DE are conserved 
separately---i.e., $\dot{\rho}_{\rm DM} +3H\rho_{\rm DCM}=0$ and $\dot{\rho}_{\rm DE}+3H\rho_{\rm DE}(1
+w_{\rm DE})=0$---is often made with almost no basis of this assumption. Then, taking advantage of the only 
remained freedom, viz., $w_{\rm DE}$, due to the unknown nature of DE, different models of DE have been put 
forward to extend the standard cosmological model since the discovery of the late 
time acceleration of the Universe. Thus, if we do not follow this two-step path to build a cosmological model, 
the fact that the nature of both DM and DE are still unknown and GR itself does not impose them to be 
conserved separately, we have, from~\cref{eqn:totconser}, $\nabla^{\mu}T_{\mu\nu}^{\rm DM}=-Q$ and
$\nabla^{\mu}T_{\mu\nu}^{\rm DE}=-Q$, namely,
\begin{align}
    \dot{\rho}_{\rm DM} +3H\rho_{\rm DM}&=Q, \label{eq:cont_eq_dm}\\
    \dot{\rho}_{\rm DE} +3H\rho_{\rm DE}(1+w_{\rm DE})&=-Q,\label{eq:cont_eq_de}
\end{align}
where we have two undetermined functions; the DE EoS parameter $w_{\rm DE}$ and the interaction kernel $Q$, 
which determines the rate and direction of the possible energy transfer between DE and DM; namely, $Q = 0$ 
implies minimal interaction (gravitational interaction only) between DM and DE, $Q>0$ implies energy transfer 
from DE to DM, and $Q<0$ implies energy transfer from DM to DE. In particular, in the case $Q = 0$ (minimal 
interaction) and $w_{\rm DE} = -1$ we have the standard  $\Lambda$CDM model. In this work, we will not impose 
any phenomenological or theoretical models for the nature of interaction between DM and DE [viz., $Q(z)$] and 
the dynamics of the DE [viz., $w_{\rm DE}$, or a corresponding $\rho_{\rm DE}(z)$], instead we will reconstruct 
these parameters, as well as some important kinematic parameters [viz., the Hubble parameter $H(z)$ and 
deceleration parameter $q(z)\equiv-1+{\rm d}H(z)^{-1}/{\rm d}t$], from observational data in a model-independent 
manner. The effects of a possible non-minimal interaction between DM and DE will be reflected on altered 
kinematics of the universe. This can be observed via the Friedmann equation~\eqref{eqn:cont}, due to the 
deviations in the evolution of the energy densities of the DM and DE from what they would have in the absence 
of a non-minimal interaction. It is in general very useful to have an idea on what corresponding minimally 
interacting (no energy exchange) DE and DM would lead to the same altered kinematics of the universe. To do so, 
we will define \textit{effective} EoS parameters for the DM and DE; $w_{\rm eff, \rm DM}$ and $w_{\rm eff, \rm DE}$,
respectively. These effective parameters are defined such that, in the absence of non-minimal interaction, they 
would lead to the same functional forms $\rho_{\rm eff, \rm DE}$ and $\rho_{\rm eff, \rm DE}$ as obtained 
through the model-independent reconstruction processes by allowing a possible non-minimal interaction. 
Accordingly, we write the following separate continuity equations for the DM and DE in terms of 
$w_{\rm eff, \rm DE}$ and $w_{\rm eff, \rm DM}$
\begin{align}
    \dot{\rho}_{\rm DM} +3H(1+w_{\rm eff, DM})\rho_{\rm DM} = 0,\\
    \dot{\rho}_{\rm DE} +3H(1+w_{\rm eff, DE})\rho_{\rm DE} = 0,
\end{align}
and then, comparing these with the continuity equations that involve the interaction kernel, i.e.,~\cref{eq:cont_eq_dm,eq:cont_eq_de}, we reach the following relation between the effective EoS parameters 
of the DM and DE and the interaction kernel:
\begin{equation}
    w_{\rm eff, \rm DM} = \frac{-Q}{3H\rho_{\rm DM}}, \quad w_{\rm eff, \rm DE} = w_{\rm DE} + \frac{Q}{3H\rho_{\rm DE}}.
\label{eq:effective_eos}
\end{equation}
It is also convenient to define a dimensionless interaction kernel parameter as follows;
\begin{equation}
    \Pi_{\rm DM} = \frac{-Q}{3H\rho_{\rm c,0}} = -\Pi_{\rm DE},
\end{equation}
where $\rho_{\rm c,0}=3H_0^2/8\pi G$ is the critical energy density of the present-day universe.
Now, let's see how $Q=Q(z)$ should behave so that we can choose appropriate priors for the reconstruction. It is 
widely accepted that, despite its problems, $\Lambda$CDM is very good at explaining most observations, so our 
efforts should not differ significantly from it, despite the model-independent nature of the reconstructions used. {For a comprehensive understanding of the impact of this interaction, perturbation analysis should also be included in our analysis. However, our focus in this study, as a proof of the concept, is on the background data and therefore we leave the full analysis including perturbations for future research. On the other hand, this does not mean that perturbations have been completely ignored here, as their effects are already reflected when choosing the prior ranges for $\Pi_{\rm DE}$.} In order to preserve the dust-like behavior of the DM and avoid significantly altering the perturbations, thus not spoiling the structure formation, one may demand that $w_{\rm eff,  DM} \sim 0$~\cite{Olivares:2007rt, 
Bolotin:2013jpa, Wang:2016lxa}, which implies $|\frac{Q}{3H\rho_{\rm DM}}|\sim 0$, then $|\frac{Q}{3H}|\ll 
\rho_{\rm DM}$. Namely, we cannot have $\frac{Q}{3H} \sim \rho_{\rm DM}>0$ otherwise the Universe would always 
remain  in the matter dominated era (viz., in the Einstein-de Sitter universe phase). Also, it is preferable  
to prevent $\frac{Q}{3H}<0$ and $|\frac{Q}{3H}| \sim \rho_{\rm DM}$ otherwise the Universe would have never 
entered the matter dominated era and the successful explanation of galaxy and large-scale structure 
formation would be spoiled.
With some algebra and using our definitions we arrive at $|\Pi_{\rm DE} | \ll \Omega_{\rm m}\frac{H^2}{H_0^2}$. Recent studies~\cite{Bernui:2023byc, Zhai:2023yny}  found that when using current cosmological data, 
the interaction could be so intense as to imply $w_{\rm eff,  DM }\sim1/3$. We will make use of these results 
as a motivation to relax the constrain on $\Pi_{\rm DE}$, so we will allow $|\Pi_{\rm DE} | \sim \Omega_{\rm m}\frac{H^2}{H_0^2}$. {These restrictions will be used as a guide when proposing the priors for the reconstruction of $\Pi_{\rm DE}(z)$ in~\cref{section:bin_gp_interpolation} and, when displaying the reconstructed $\Pi_{\rm DE}$, we will plot the curve $\Omega_{\rm m}H^2/H_0^2$ as a reference.}

By using the dimensionless interaction kernel together with the chain rule and $\rho_{\rm c,0}$, we can express 
\cref{eq:cont_eq_dm,eq:cont_eq_de} as 
\begin{subequations}\label{subeqs:cont}
     \begin{align}
       \frac{{\rm d}(\rho_{\rm DM}/\rho_{\rm c,0})}{{\rm d}z} &= \frac{3}{1+z}\bigg ( 
       \frac{\rho_{\rm DM}}{\rho_{\rm c,0}} + \Pi_{\rm DM} \bigg ), \label{subeq:cont_dens_param_dm}\\
     \frac{{\rm d}(\rho_{\rm DE}/\rho_{\rm c,0})}{{\rm d}z} &= \frac{3}{1+z}\bigg[(1
     +w_{\rm DE})\frac{\rho_{\rm DE}}{\rho_{\rm c,0}} + \Pi_{\rm DE} \bigg], \label{subeq:cont_dens_param_de}
     \end{align}
\end{subequations}
respectively. These continuity equations are then solved numerically and used to express our Friedmann equation,
i.e., $H(z)$. The continuity equations for radiation and baryonic matter do not change, so we have, 
assuming a spatially flat universe:
\begin{equation}
    \frac{H^2(z)}{H_0^2} =  \Omega_{b,0}(1+z)^3 +  \frac{\rho_{\rm DM}(z)}{\rho_{\rm c,0}}+\frac{\rho_{\rm DE}(z)}{\rho_{\rm c,0}},
\end{equation}
where we have neglected radiation, as it is well negligible in the post-recombination universe.

In~\cite{vonMarttens:2019ixw} it was demonstrated that an equivalence between dynamical DE (through a 
dynamical EoS parameter) and an interacting DE-DM model (with a constant EoS parameter) exists at the 
background level. To avoid this and to maintain as little bias as possible regarding the underlying 
possible functional form of the dimensionless interaction kernel parameter $\Pi_{\rm DE}(z)$, our 
reconstruction efforts will be aimed mainly towards the interaction kernel but letting the EoS parameter 
to be a variable single bin $w_0$. Reconstructing both functions with model independent approaches, 
with a large number of extra parameters at the same time, could lead to a lot of degeneracies with
\cref{eq:effective_eos}, but it may be worth to do it in future works.

Finally, for the sake of comparison we will also plot Solano's dimensionless interaction function~\cite{Solano:2012zw}:
\begin{equation}
    I_Q(z) = \frac{Q(z)}{\rho_{\rm c,0} H(z) (1+z)^3},
\end{equation}
which is proposed as a way to better visualize the interaction kernel and its defining characteristics.

\section{BINNED AND GAUSSIAN PROCESS INTERPOLATIONS}\label{section:bin_gp_interpolation}

One of the reconstruction methods considered in this paper, to describe $f(z)$,  
consists in using a set of step functions connected via hyperbolic tangents to maintain smooth continuity. The function to reconstruct then takes the 
following form:
\begin{equation}
    f(z)= f_1+\sum^{N-1}_{i=1}\frac{f_{i+1} - f_i}{2}\bigg[1+\tanh{\Big(\frac{z-z_i}{\xi}\Big)} \bigg],
\label{bin_equation}
\end{equation}
where $N$ is the number of bins, $f_i$ the amplitude of the bin value, $z_i$ the position where the bin begins in the $z$ axis and $\xi$ the smoothness parameter, set to $\xi = 0.15$ in this work.

The other approach is an interpolation by using a Gaussian Process (GP). A GP is the 
generalization of a Gaussian distribution, that is, in every position $x$, $f(x)$ is a random variable. 
It is characterized by a mean function $\mu(x)$ and a covariance $\sigma^2K(x,x')$, where $\sigma^2$ is 
the variance and $K(x,x')$ the kernel representing the correlation of $f$ between two different positions 
$f(x)$ and $f(x')$. For an arbitrary amount of positions $x_1,..,x_n$ then we have a multivariate Gaussian
distribution
\begin{equation}
    \Bar{f}=[f(x_1),..,f(x_n)]~\Bar{N}(\Bar{\mu}, \sigma^2K(\Bar{x},\Bar{x'})),
\end{equation}
where $\Bar{\mu}=[\mu(x_1),...,\mu(x_n)]$, and
\begin{equation}
K(\Bar{x},\Bar{x}') = 
\begin{pmatrix}
K(x_1,x_1) & K(x_1,x_2) & \cdots & K(x_1,x_n) \\
K(x_2,x_1) & K(x_2,x_2) & \cdots & K(x_2,x_n) \\
\vdots  & \vdots  & \ddots & \vdots  \\
K(x_n,x_1) & K(x_n,x_2) & \cdots & K(x_n,x_n) 
\end{pmatrix}.
\end{equation}

The Kernel we will use in this work is the Radial Basis Function (RBF)
\begin{equation}
    K(x,x')= {\rm exp}\big[-\theta(x-x')^2 \big],
\end{equation}
where the parameter $\theta$ tells us how strong is the correlation. This kernel has the advantage of 
minimizing the degeneracies created due to a high number of hyperparameters since it only has $\theta$, 
it is isotropic if we choose $x=z$ being $z$ the cosmological redshift, and it is also infinitely 
differentiable. 

\begin{figure}[t!]
    \begin{center}
     \includegraphics[ width=9.cm, height=5.cm]{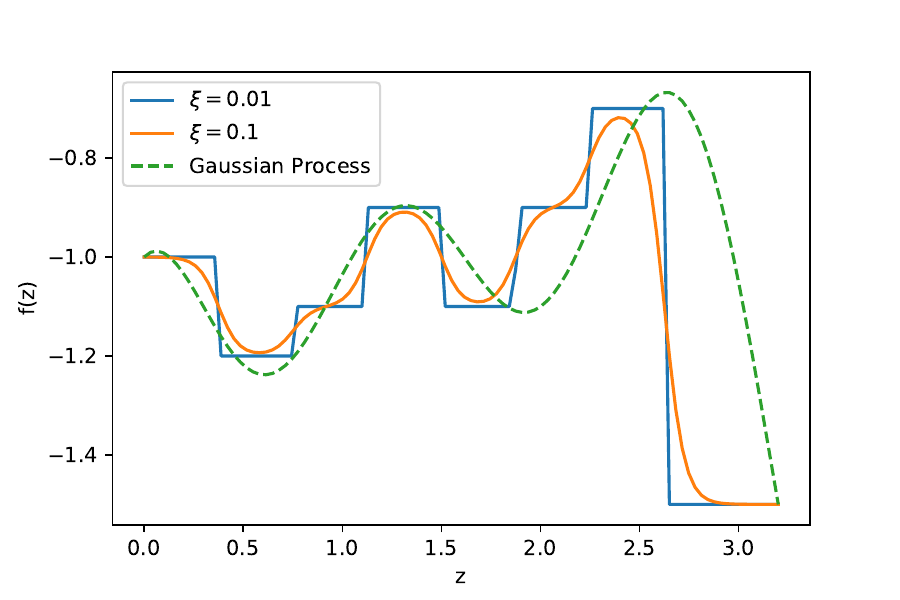}
    \end{center}
    \caption{Comparison between a Gaussian Process for interpolation and a step function approach. 
    The influence of the smoothness parameter $\xi$ in the Binning scheme is also shown.
    }\label{fig:comparison_tanh_gp}
\end{figure}

Analogous to the binning approach, the GP will be used as an interpolation between nodes in order 
to have a model-independent reconstruction in a similar fashion as the reconstruction performed 
in~\cite{Gerardi:2019obr}. This method yields to slightly different results as seen in~\cref{fig:comparison_tanh_gp}. 
We will have node values located at $z_i$ to described 
$\Pi_{\rm DE}(z_i)$. The $z_i$ values remained fixed, so the free parameters for our interaction 
kernel would be the amplitudes $\Pi_{\rm DE}(z_i)=\Pi_i$. 

{In the present work, and without loss of generality, for the reconstruction of $\Pi_{\rm DE}$ we will utilize five amplitudes, evenly located across the interval of $0.0 \leq z \leq 3.0$. This choice implies that each amplitude encompasses a redshift interval of 0.6 when using bins. Alternatively, when utilizing GP the positions of the amplitudes $\Pi_i$ are located in the following positions $[0.0, 0.75, 1.5, 2.25, 3.0]$. }

\section{DATASETS AND METHODOLOGY}\label{section:data}

In this work, we will use the collection of cosmic chronometers~\cite{jimenez2003constraints, 
simon2005constraints, stern2010cosmic, moresco2012new, zhang2014four, moresco2015raising, moresco20166} 
(we will refer to this dataset as H), which can be found within the repository~\cite{hz}.
%
We also make use of the full catalogue of supernovae from the Pantheon+ SN Ia sample, covering a 
redshift range of $0<z\lesssim2.26$~\cite{Scolnic:2021amr} (we will refer to this dataset as SN). 
The full covariance matrix associated is comprised of a statistical and a systematic part, and along 
with the data, they are provided in the repository~\cite{pantheon_data}.
Finally we also employ the BAO datasets, containing the SDSS Galaxy Consensus, quasars and Lyman-$\alpha$ 
forests~\cite{eBOSS:2020yzd}. The sound horizon is calibrated by using BBN~\cite{Cooke:2013cba}.
For a more detailed description of the datasets refer to~\cite{Escamilla:2021uoj}. We will call 
this dataset BAO.\\
%

\begin{table}[t!]
\caption{Jeffreys' scale for model selection with the logarithm of the Bayes' factor. 
Using the convention from~\cite{Trotta:2008qt}.}
\footnotesize
\scalebox{1}{%
\begin{tabular}{cccc} 
\cline{1-4}\noalign{\smallskip}
 \vspace{0.15cm}

 $\ln{B_{12}}$ & Odds  &   Probability  &  Strength of evidence \\
\hline
 
\hline
\vspace{0.15cm}
$<$ 1.0 & $<$3:1 & $<$0.75 & Inconclusive  \\
\vspace{0.15cm}
1.0 &  $\sim$3:1 & 0.750 & Weak evidence  \\
\vspace{0.15cm}
2.5 & $\sim$12:1  &  0.923  & Moderate evidence  \\
\vspace{0.15cm}
5.0 & $\sim$150:1 &  0.993  & Strong evidence  \\

\hline
\hline
\end{tabular}}
\label{jeffreys}
\end{table}

To find the best-fit values for the free parameters of our model, we use a modified version of 
the Bayesian inference code, called SimpleMC \cite{simplemc, aubourg2015cosmological}, used for 
computing expansion rates 
and distances from the Friedmann equation. For a model $i$ we have computed its Bayesian evidence $E_i$, 
and to compare two different models (1 and 2) we make use of the Bayes' factor $B_{1,2} = E_1/E_2$, 
specifically its natural logarithm. When used in tandem with the empirical Jeffreys' scale,~\cref{jeffreys}, 
we can have a better notion of the alternative models' performance. To evaluate the fitness of our 
reconstructions (with respect to $\Lambda$CDM) we will make use of the $-2\ln \mathcal{L_{\rm max}}$ 
of each model, where $\mathcal{L_{\rm max}}$ is the maximum likelihood obtained (in the Bayesian sense).\footnote{See~\cite{Padilla:2019mgi} for a cosmological Bayesian inference review.} 
The SimpleMC code includes the \texttt{dynesty} library~\cite{speagle2020dynesty},
a nested sampling algorithm used to compute the Bayesian evidence.  The number of live-points were 
selected using the general rule $50 \times ndim$~\cite{dynestyy}, where $ndim$ is the number of parameters 
to be sampled. 
The flat priors used for the base parameters are: $\Omega_{\rm m} = [0.1, 0.5]$ for the matter 
density parameter, $\Omega_{\rm b} h^2 = [0.02, 0.025]$ for the physical baryon density, $h = [0.4, 0.9]$ 
for the dimensionless Hubble constant $h=H_0/100\, {\rm km\,s}^{-1}{\rm Mpc}^{-1}$. For comparison we 
include the $w$CDM model \wde{}$ = w_c$, the Chevallier-Polarski-Linder (CPL) 
EoS parameter \wde{}$= w_0 + w_a\frac{z}{1+z}$~\cite{Chevallier} and the sign-switch interaction 
kernel (SSIK) $Q=3\sigma H(\rho_{\rm DE}-\alpha\rho_{\rm DM})$~\cite{Sun:2010vz}. Their free parameters 
being $w_c$, $w_0$ and $w_a$ for $w$CDM and CPL; for SSIK $w_0 = [-2.0, 0.0]$, $\sigma = [0.0, 1.0]$ 
and $\alpha= [0, 4]$.
The flat prior for $w_c$, and $w_0$, is the same $[-2.0,0.0]$ and the flat prior used for $w_a$ is 
$[-2.0, 2.0]$. For the reconstruction of $\Pi_{\rm DE}(z)$ we recall that $|\Pi_{\rm DE} | 
\ll \Omega_{\rm m}(z)E^2(z)$, with $E(z)\equiv \frac{H(z)}{H_0}$, also that generally $\Omega_{\rm m}(0) 
\approx 0.3$, and as we move from late-times to early-times this value only grows. 
We will use $|\Pi_{\rm DE}| \sim \Omega_{\rm m}(z)E^2(z)$ as a loose guide as mentioned before to choose our priors, we 
have then $\Pi_{\rm DE,i} = [-2.0,2.0]$ when $z<1.0$ and $\Pi_{\rm DE,i} = [-15.0,15.0]$ when $z>1.0$ .
Regarding the EoS parameter we either fix it to a cosmological 
constant $w_0 = -1$ or let it vary as a free parameter $w_0 = [-2.0, 0.0]$.

\section{RESULTS}\label{section:results}

\begin{figure*}[t!]
    \centering
    \makebox[11cm][c]{
    \includegraphics[trim = 5mm  0mm 25mm 0mm, clip, width=5.cm, height=4.5cm]{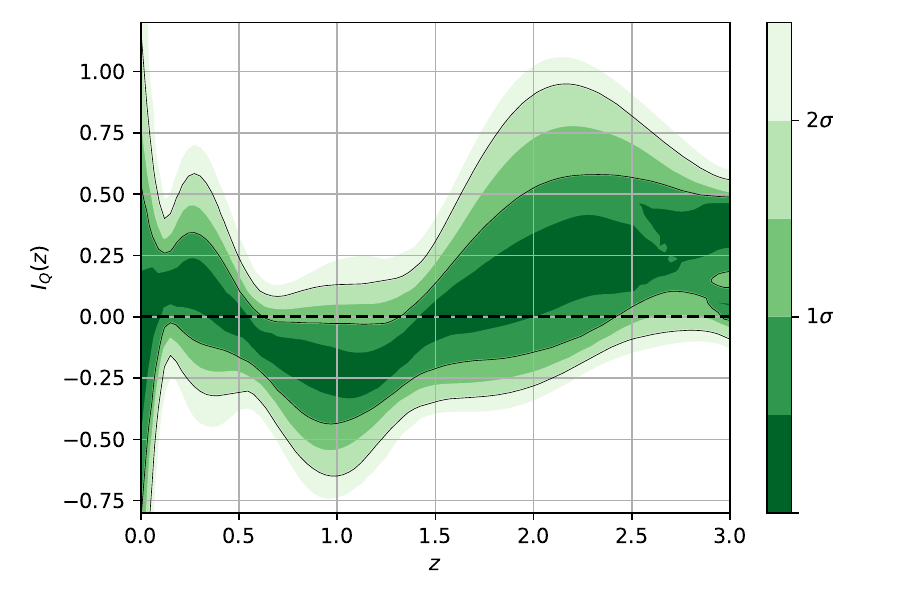}
    \includegraphics[trim = 5mm  0mm 25mm 0mm, clip, width=5.cm, height=4.5cm]{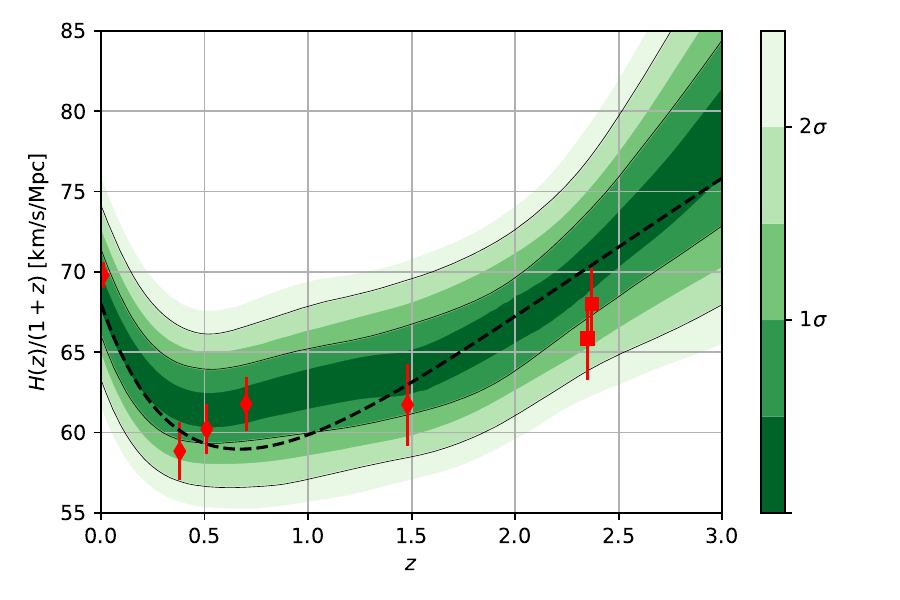}
    \includegraphics[trim = 5mm  0mm 25mm 0mm, clip, width=5.cm, height=4.5cm]{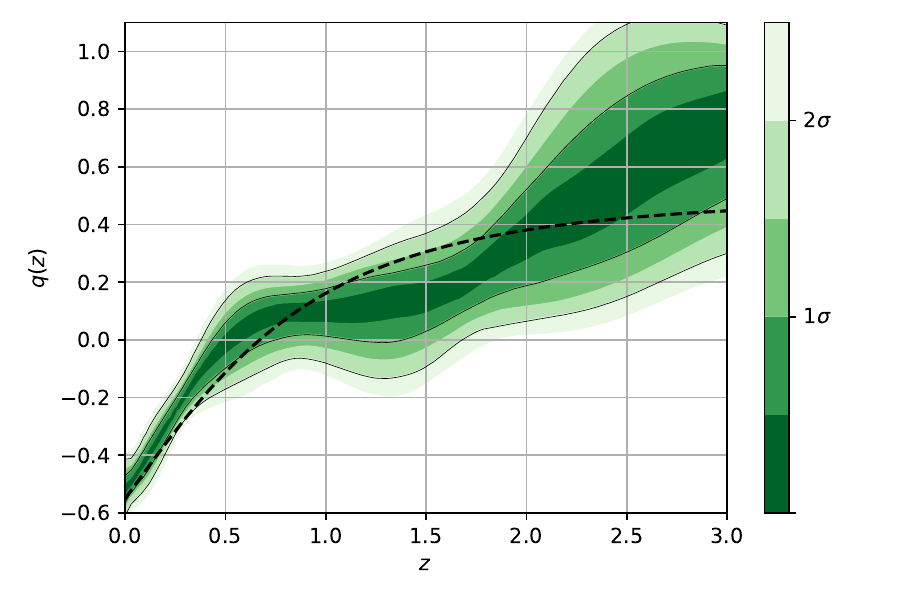}
    \includegraphics[trim = 5mm  0mm 5mm 0mm, clip, width=5.5cm, height=4.5cm]{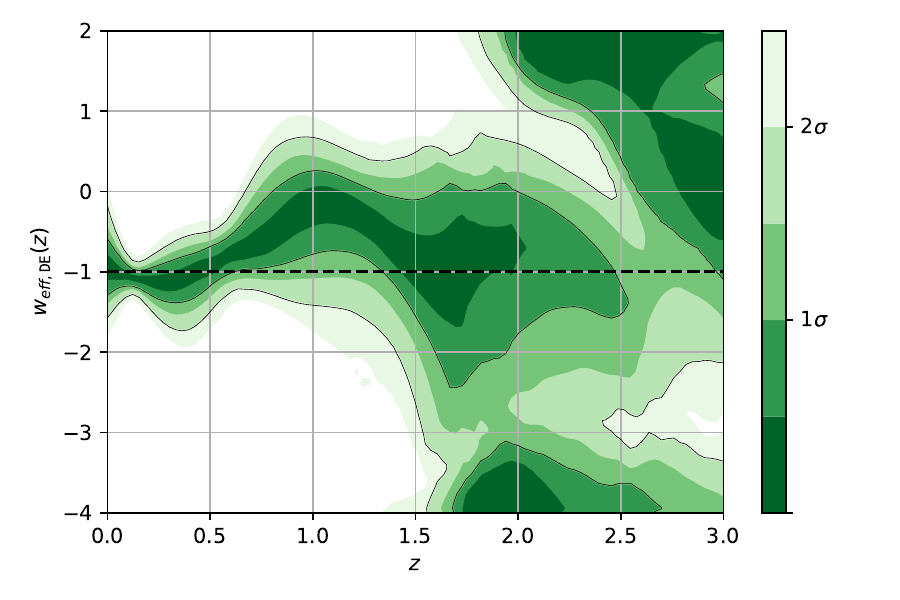}
    }
    \makebox[11cm][c]{
    \includegraphics[trim = 5mm  0mm 25mm 0mm, clip, width=5.cm, height=4.5cm]{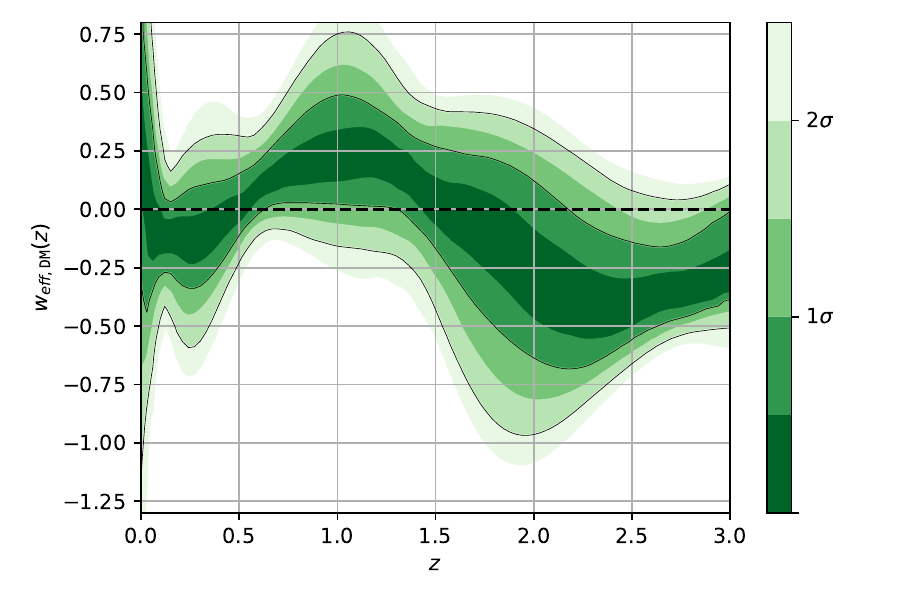}
    \includegraphics[trim = 5mm  0mm 25mm 0mm, clip, width=5.cm, height=4.5cm]{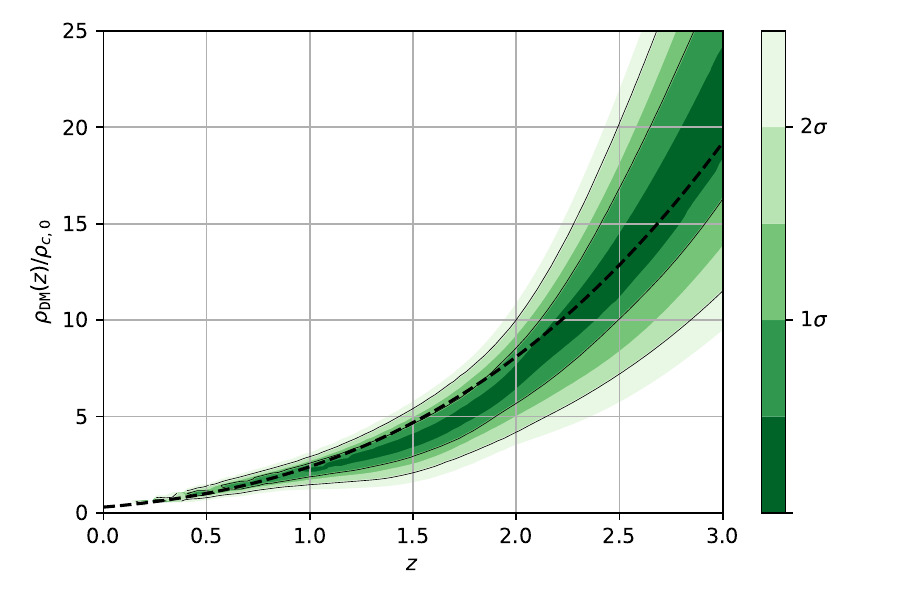}
    \includegraphics[trim = 5mm  0mm 25mm 0mm, clip, width=5.cm, height=4.5cm]{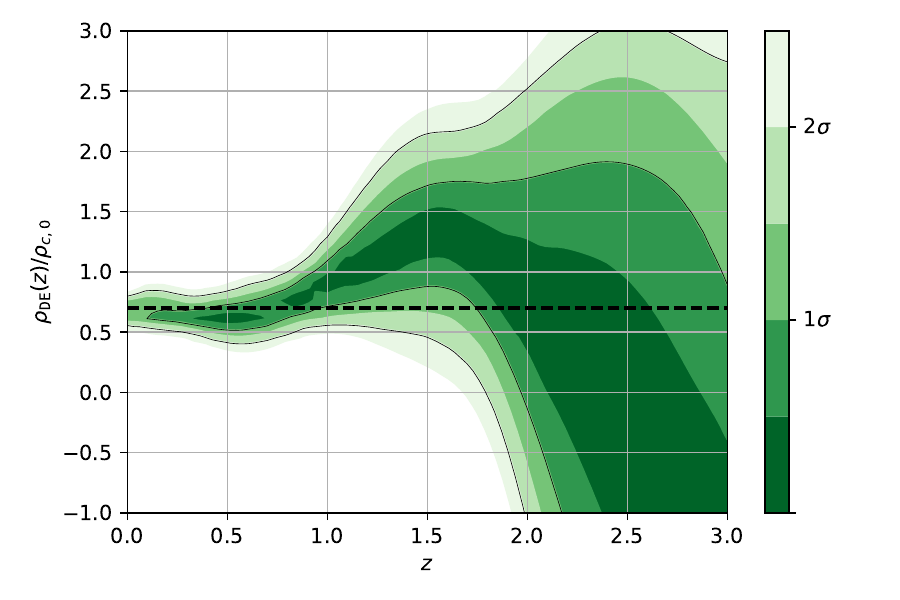}
    \includegraphics[trim = 5mm  0mm 5mm 0mm, clip, width=5.5cm, height=4.5cm]{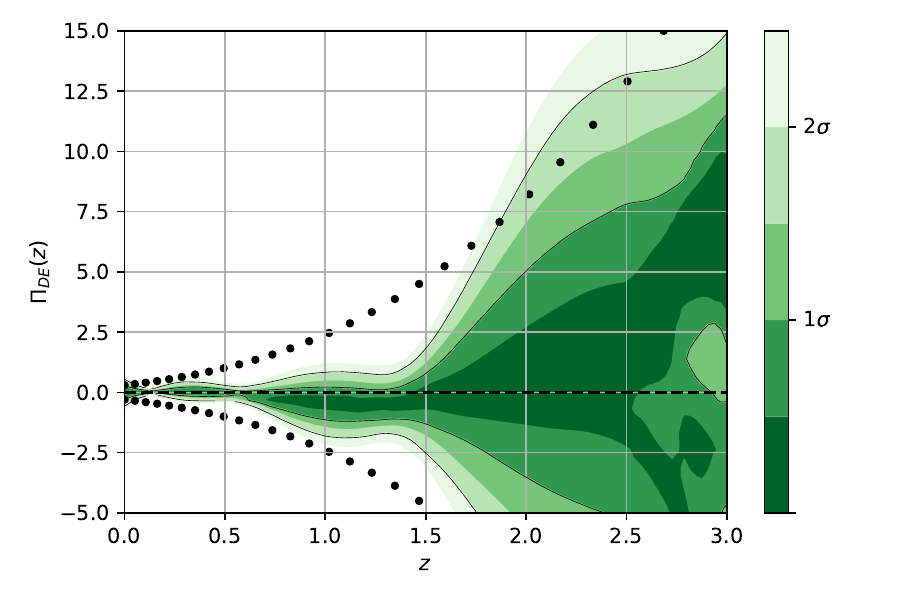}
    }
    \caption{Functional posterior probability of the reconstruction by using a Gaussian Process and $w=-1$. 
    The probability as normalised in each slice of constant $z$, with colour scale in confidence interval values (see color bar at the right). 
    The 68\% ($1\sigma$) and 95\% ($2\sigma$) confidence intervals are plotted as black lines. From left to right 
    in the upper part: the reescalation function $I_Q(z)$, the Hubble Parameter, the deceleration parameter and 
    the effective EoS parameter for DE. In the lower part: the effective EoS parameter for DM, the density for 
    DM and DE respectively and the dimensionless interaction kernel $\Pi_{\rm DE}$. The dashed black line corresponds to 
    the standard $\Lambda$CDM values {and the dotted line in the $\Pi_{\rm DE}(z)$ plot corresponds to the $\Omega_{\rm m}H(z)^2/H_0^2$ curve.}
}\label{fig:derived_gp_nowide}
\end{figure*}

\begin{figure*}[t!]
    \centering
    \makebox[11cm][c]{
    \includegraphics[trim = 5mm  0mm 25mm 0mm, clip, width=5.cm, height=4.5cm]{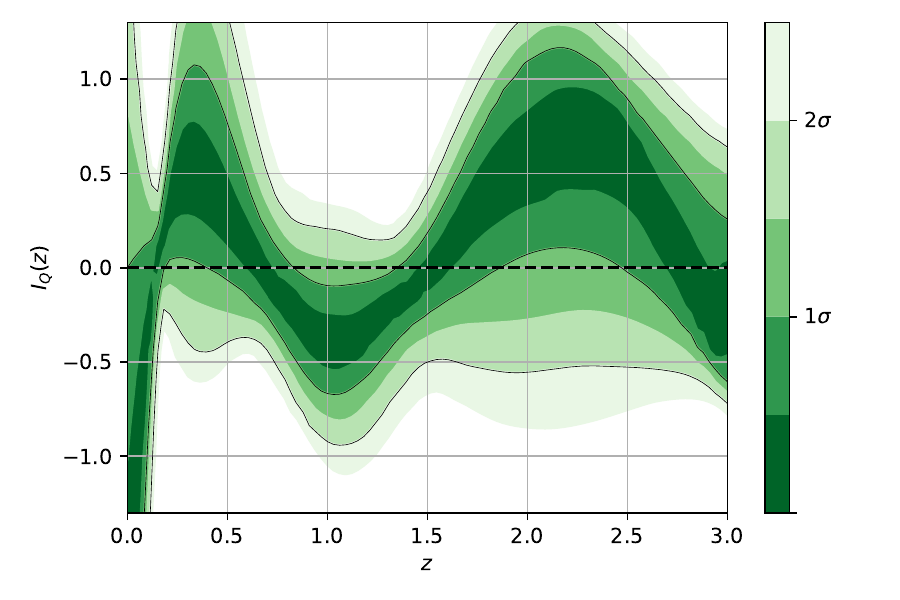}
    \includegraphics[trim = 5mm  0mm 25mm 0mm, clip, width=5.cm, height=4.5cm]{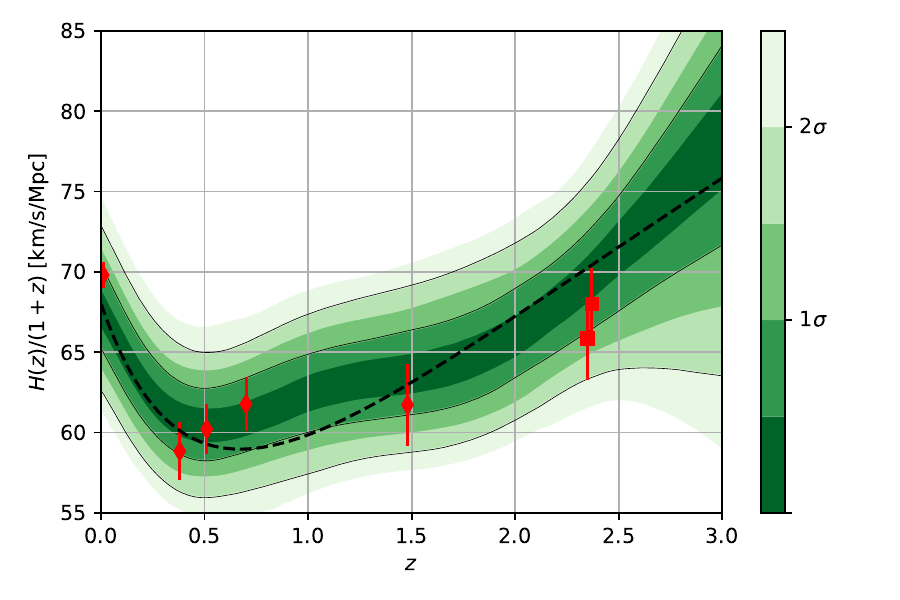}
    \includegraphics[trim = 5mm  0mm 25mm 0mm, clip, width=5.cm, height=4.5cm]{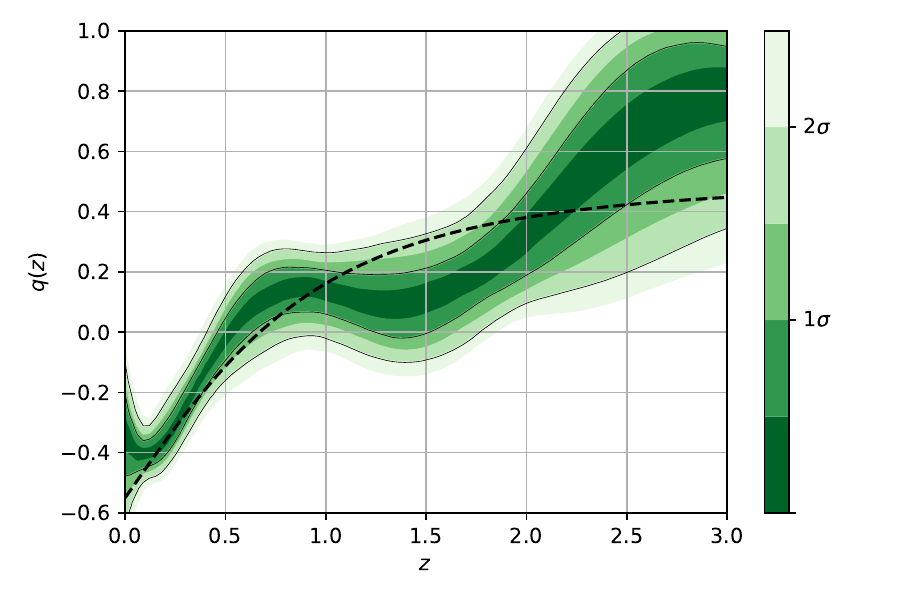}
    \includegraphics[trim = 5mm  0mm 5mm 0mm, clip, width=5.5cm, height=4.5cm]{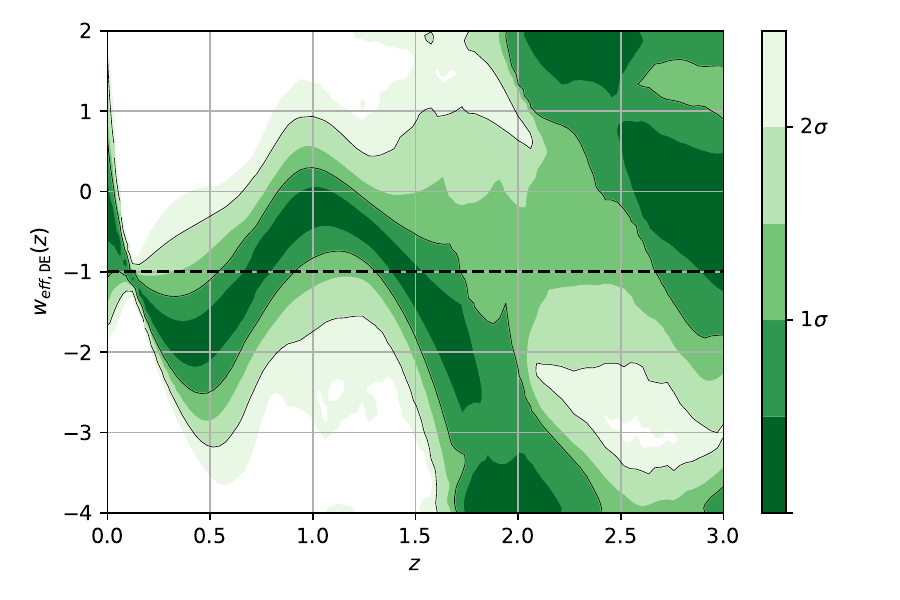}
    }
    \makebox[11cm][c]{
    \includegraphics[trim = 5mm  0mm 25mm 0mm, clip, width=5.cm, height=4.5cm]{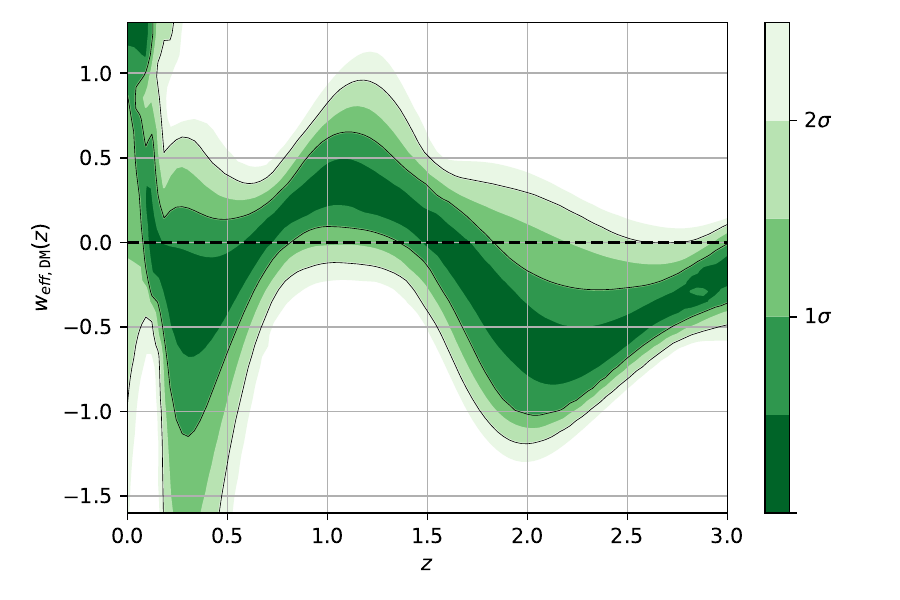}
    \includegraphics[trim = 5mm  0mm 25mm 0mm, clip, width=5.cm, height=4.5cm]{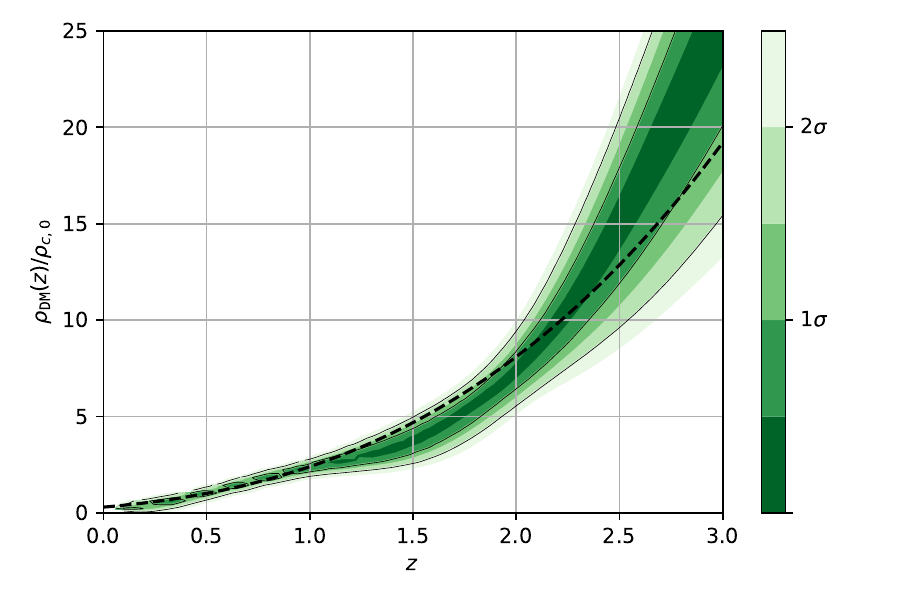}
    \includegraphics[trim = 5mm  0mm 25mm 0mm, clip, width=5.cm, height=4.5cm]{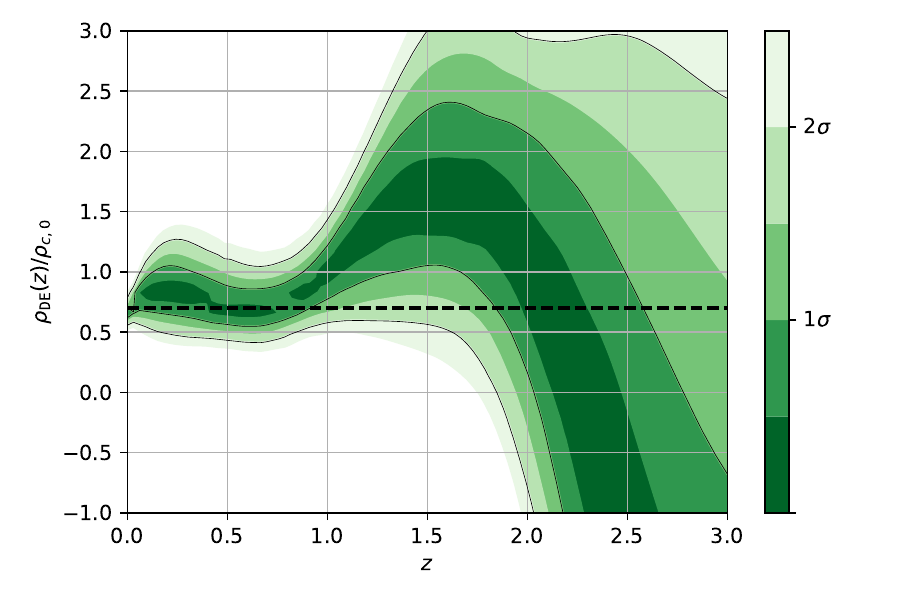}
    \includegraphics[trim = 5mm  0mm 5mm 0mm, clip, width=5.5cm, height=4.5cm]{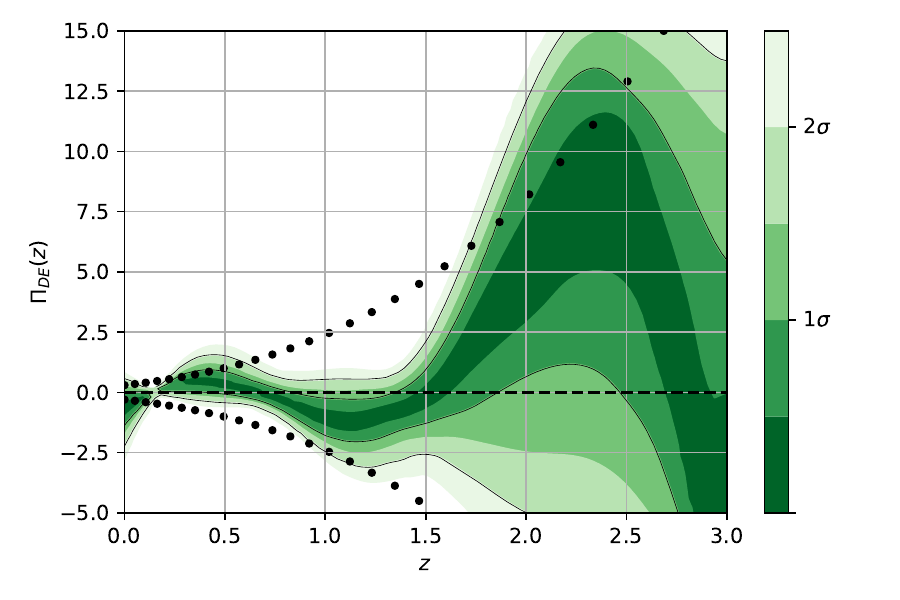}
    }
    \caption{Functional posterior probability of the reconstruction by using a Gaussian Process and $w=w_0$. 
    The probability as normalised in each slice of constant $z$, with colour scale in confidence interval values. 
    The 68\% ($1\sigma$) and 95\% ($2\sigma$) confidence intervals are plotted as black lines. From left to 
    right in the upper part: the reescalation function $I_Q(z)$, the Hubble Parameter, the deceleration 
    parameter and the effective EoS parameter for DE. In the lower part: the effective EoS parameter for DM, 
    the density for DM and DE respectively and the dimensionless interaction kernel $\Pi_{\rm DE}$. 
    The dashed black line corresponds to the standard $\Lambda$CDM values {and the dotted line in the $\Pi_{\rm DE}(z)$ plot corresponds to the $\Omega_{\rm m}H(z)^2/H_0^2$ curve.}
}\label{fig:derived_gp_wide}
\end{figure*}

\begin{figure*}[t!]
    \centering
    \makebox[11cm][c]{
    \includegraphics[trim = 5mm  0mm 25mm 0mm, clip, width=5.cm, height=4.5cm]{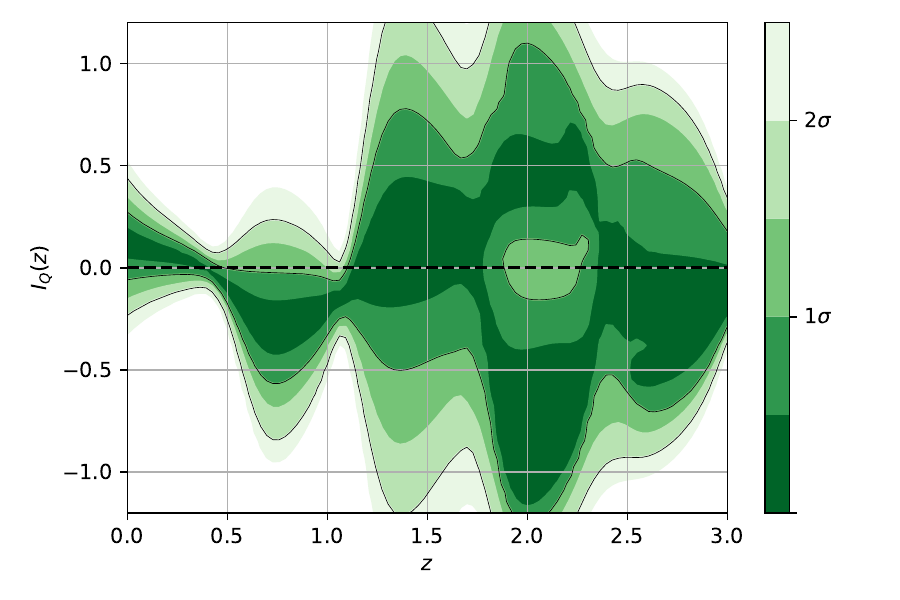}
    \includegraphics[trim = 5mm  0mm 25mm 0mm, clip, width=5.cm, height=4.5cm]{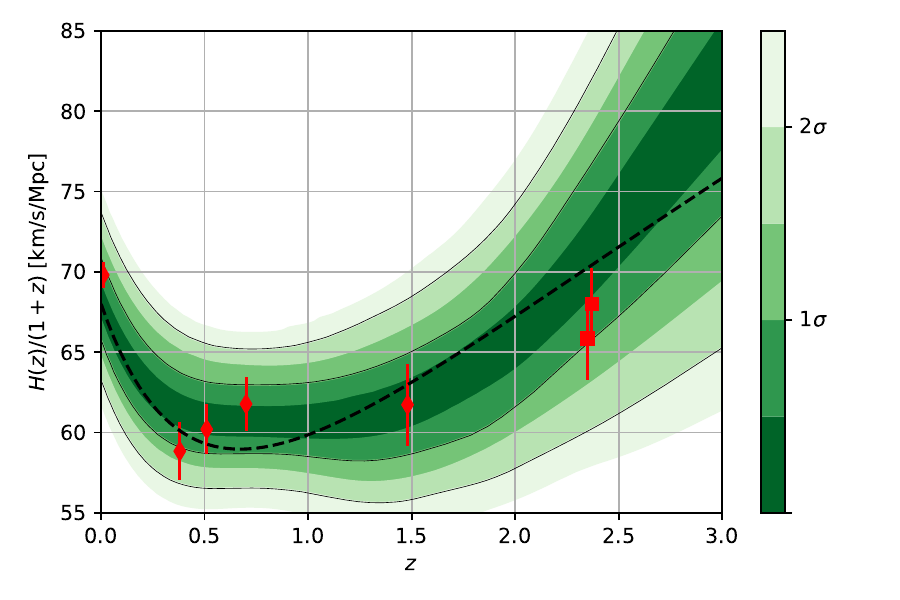}
    \includegraphics[trim = 5mm  0mm 25mm 0mm, clip, width=5.cm, height=4.5cm]{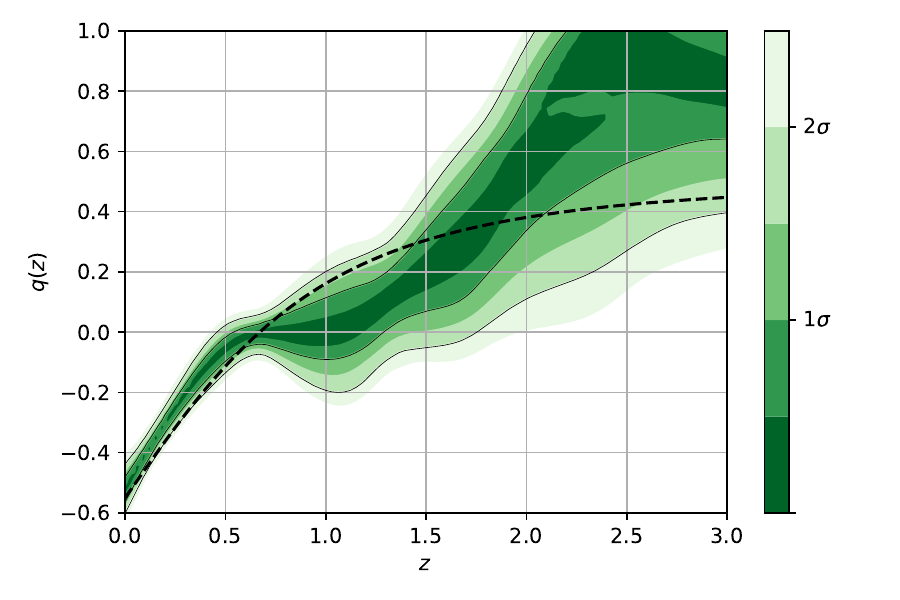}
    \includegraphics[trim = 5mm  0mm 5mm 0mm, clip, width=5.5cm, height=4.5cm]{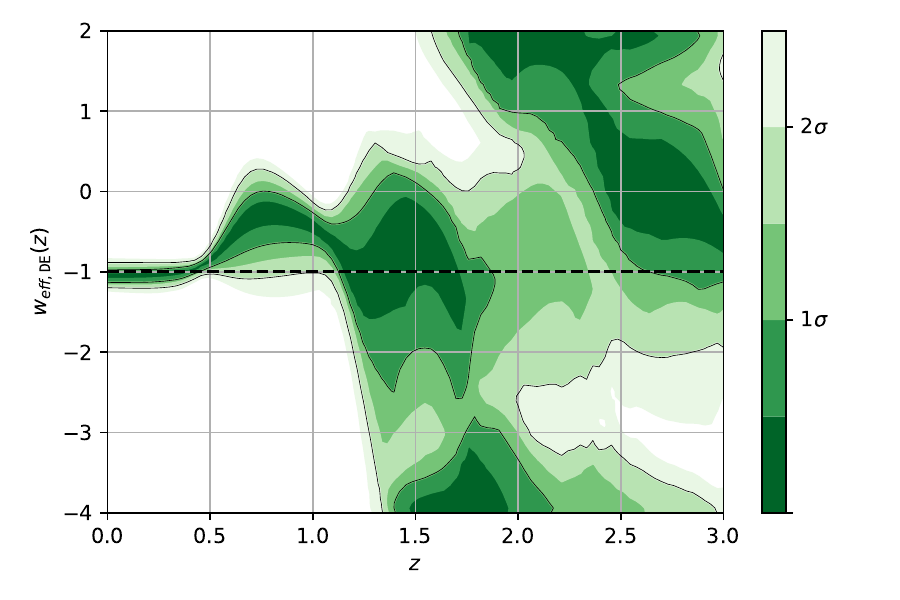}
    }
    \makebox[11cm][c]{
    \includegraphics[trim = 5mm  0mm 25mm 0mm, clip, width=5.cm, height=4.5cm]{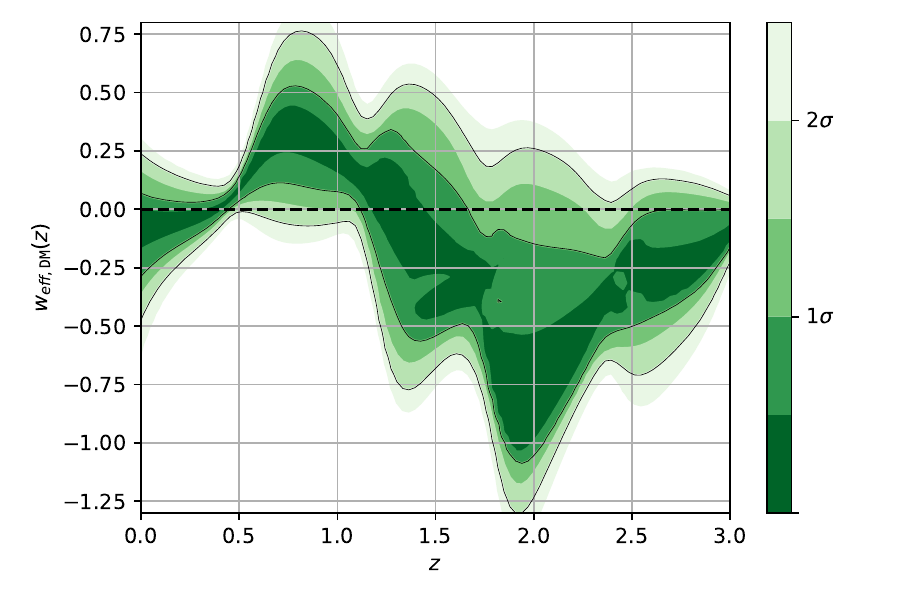}
    \includegraphics[trim = 5mm  0mm 25mm 0mm, clip, width=5.cm, height=4.5cm]{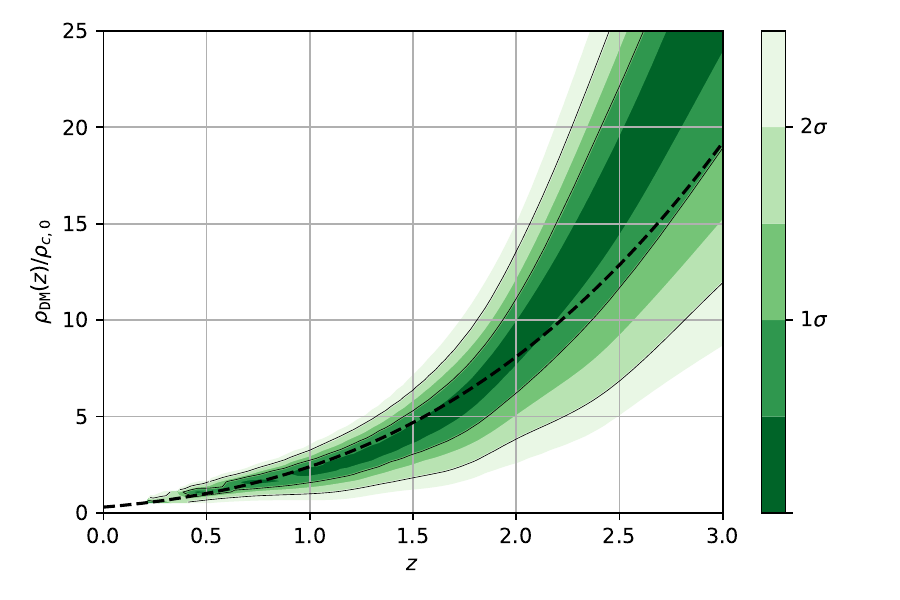}
    \includegraphics[trim = 5mm  0mm 25mm 0mm, clip, width=5.cm, height=4.5cm]{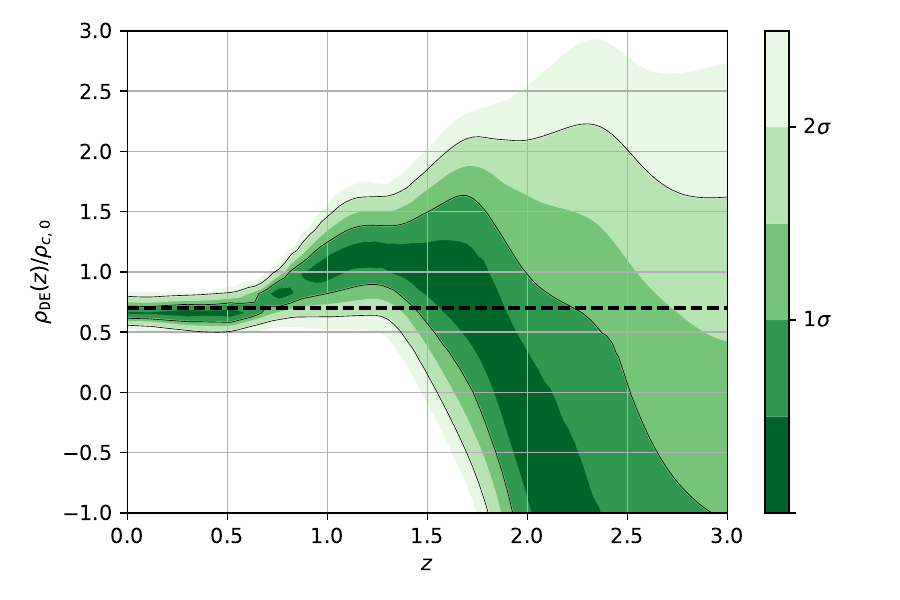}
    \includegraphics[trim = 5mm  0mm 5mm 0mm, clip, width=5.5cm, height=4.5cm]{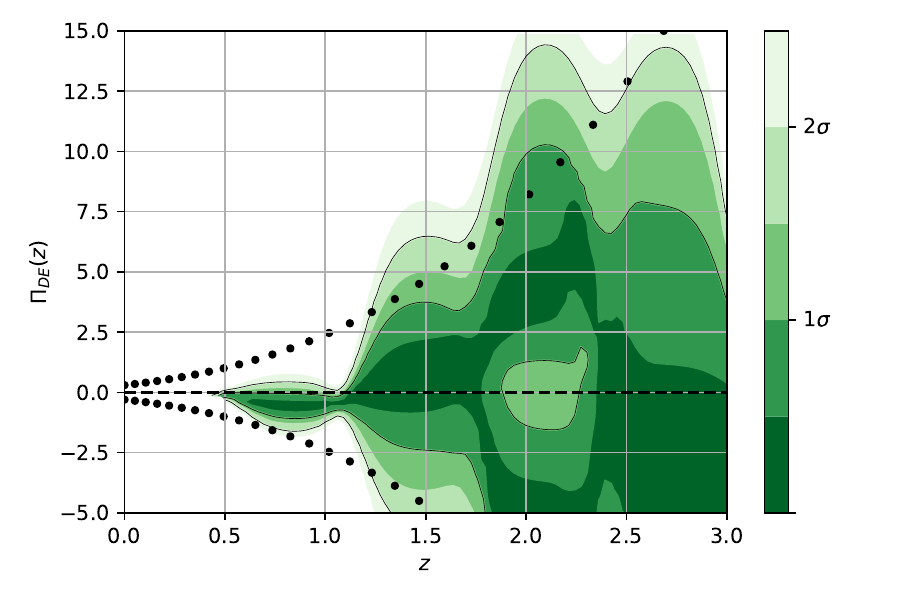}
    }
    \caption{Functional posterior probability of the reconstruction by using a Binning scheme and $w=-1$. 
    The probability as normalised in each slice of constant $z$, with colour scale in confidence interval 
    values. The 68\% ($1\sigma$) and 95\% ($2\sigma$) confidence intervals are plotted as black lines. 
    From left to right in the upper part: the reescalation function $I_Q(z)$, the Hubble Parameter, 
    the deceleration parameter and the effective EoS parameter for DE. In the lower part: the effective 
    EoS parameter for DM, the density for DM and DE respectively and the dimensionless interaction 
    kernel $\Pi_{\rm DE}$. The dashed black line corresponds to the standard $\Lambda$CDM values {and the dotted line in the $\Pi_{\rm DE}(z)$ plot corresponds to the $\Omega_{\rm m}H(z)^2/H_0^2$ curve.}
}\label{fig:derived_tanh_nowide}
\end{figure*}

\begin{figure*}[t!]
    \centering
    \makebox[11cm][c]{
    \includegraphics[trim = 5mm  0mm 25mm 0mm, clip, width=5.cm, height=4.5cm]{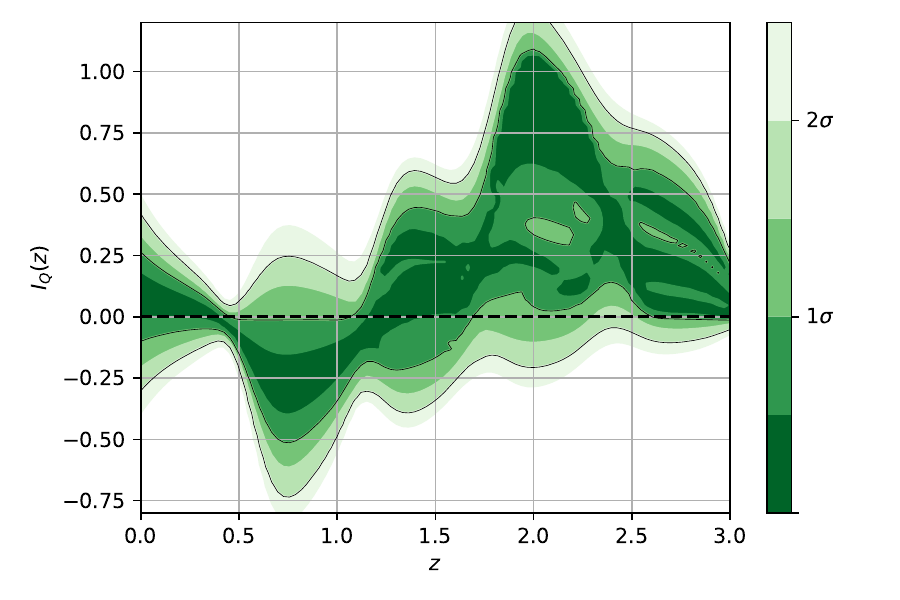}
    \includegraphics[trim = 5mm  0mm 25mm 0mm, clip, width=5.cm, height=4.5cm]{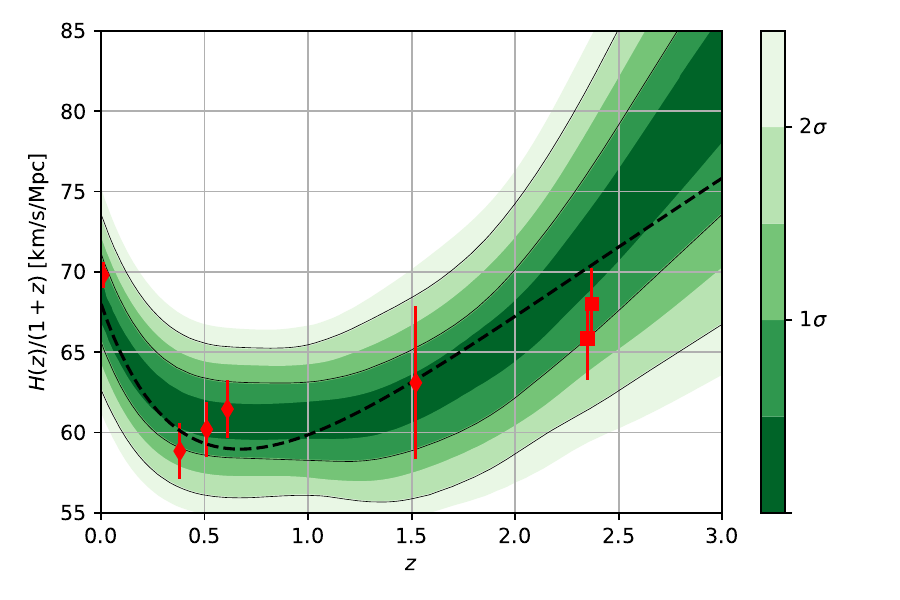}
    \includegraphics[trim = 5mm  0mm 25mm 0mm, clip, width=5.cm, height=4.5cm]{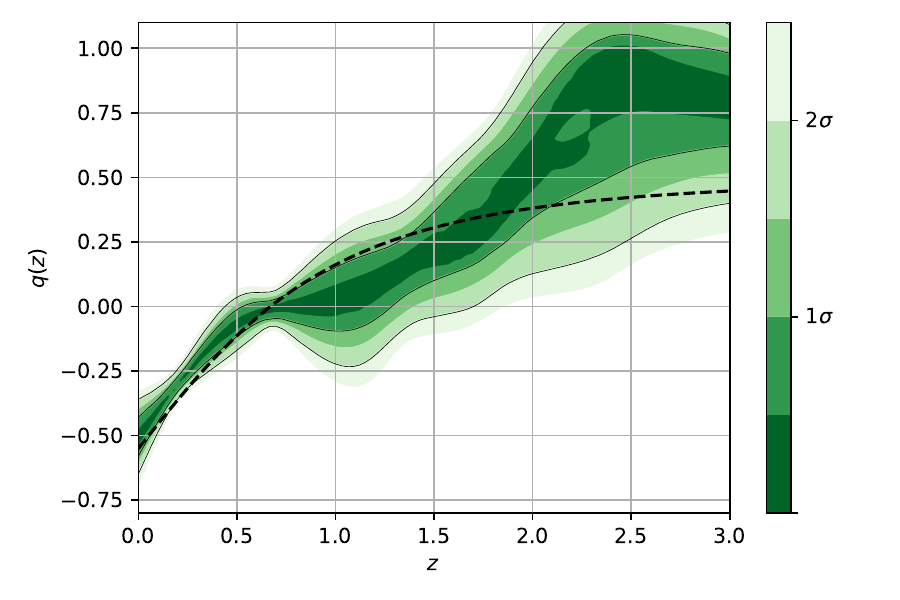}
    \includegraphics[trim = 5mm  0mm 5mm 0mm, clip, width=5.5cm, height=4.5cm]{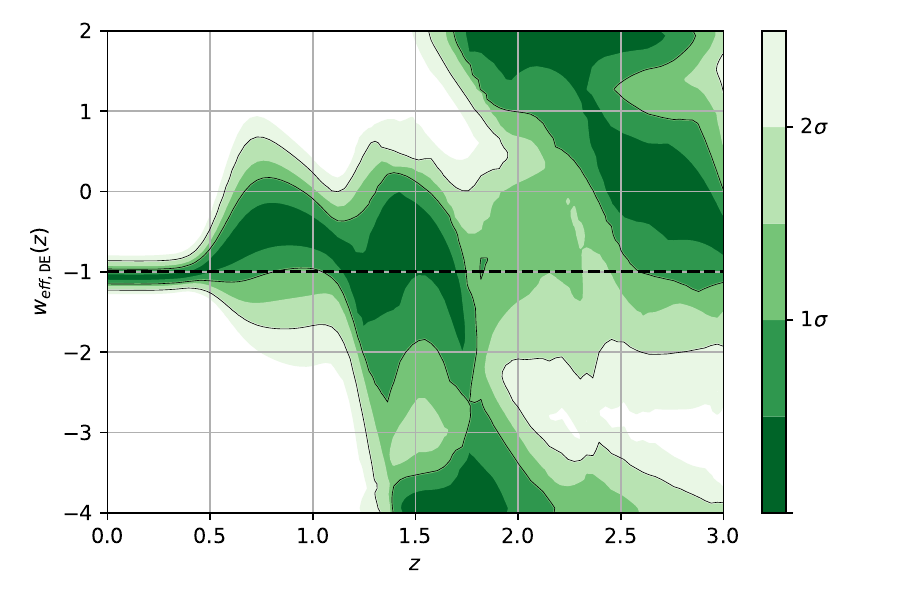}
    }
    \makebox[11cm][c]{
    \includegraphics[trim = 5mm  0mm 25mm 0mm, clip, width=5.cm, height=4.5cm]{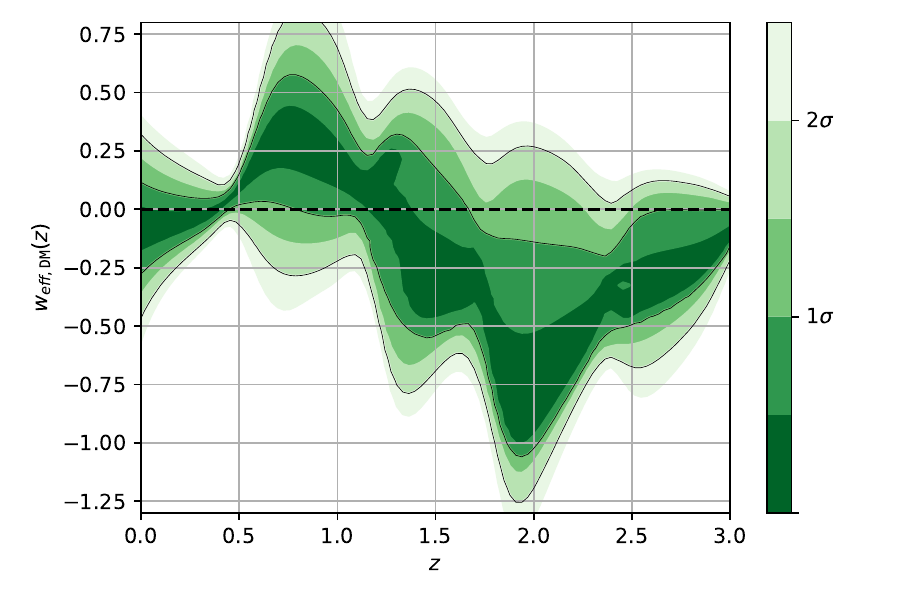}
    \includegraphics[trim = 5mm  0mm 25mm 0mm, clip, width=5.cm, height=4.5cm]{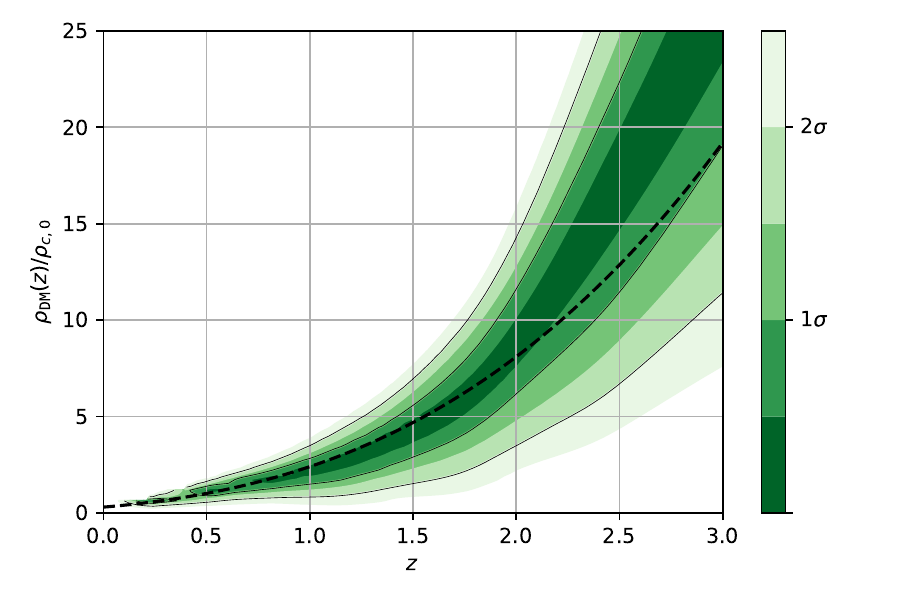}
    \includegraphics[trim = 5mm  0mm 25mm 0mm, clip, width=5.cm, height=4.5cm]{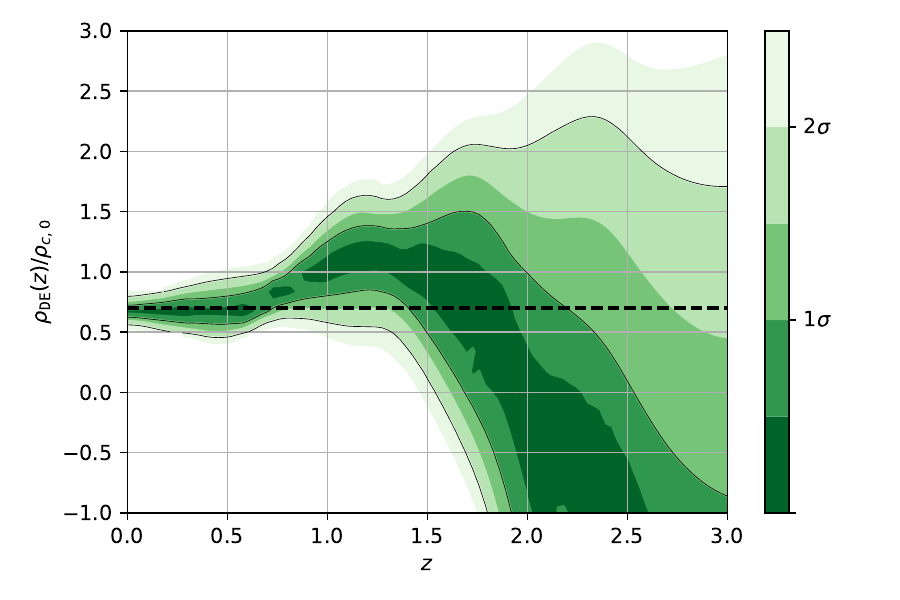}
    \includegraphics[trim = 5mm  0mm 5mm 0mm, clip, width=5.5cm, height=4.5cm]{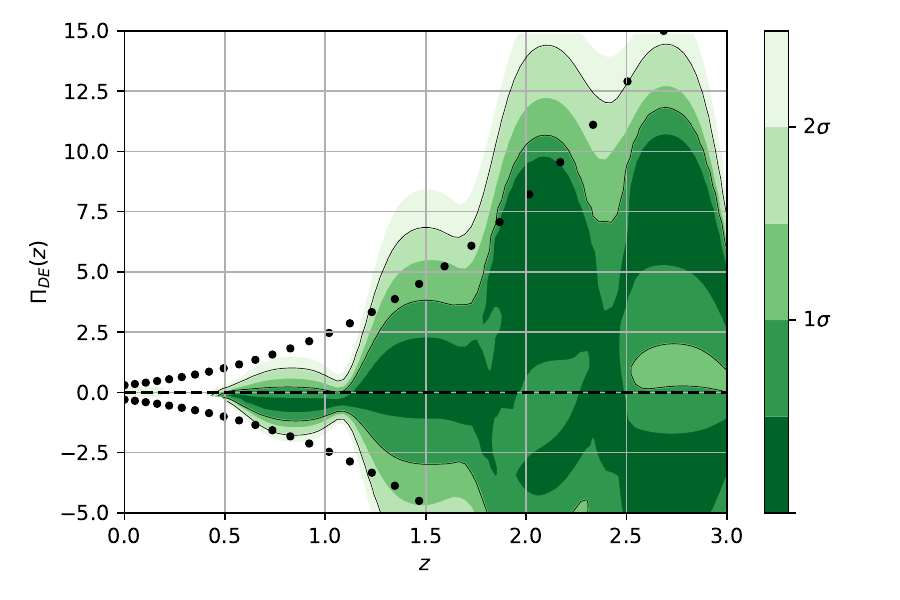}
    }
    \caption{Functional posterior probability of the reconstruction by using a Binning scheme and $w=w_0$. 
    The probability as normalised in each slice of constant $z$, with colour scale in confidence interval 
    values. The 68\% ($1\sigma$) and 95\% ($2\sigma$) confidence intervals are plotted as black lines. 
    From left to right in the upper part: the reescalation function $I_Q(z)$, the Hubble Parameter, 
    the deceleration parameter and the effective EoS parameter for DE. In the lower part: the effective 
    EoS parameter for DM, the density for DM and DE respectively and the dimensionless interaction 
    kernel $\Pi_{\rm DE}$. The dashed black line corresponds to the standard $\Lambda$CDM values {and the dotted line in the $\Pi_{\rm DE}(z)$ plot corresponds to the $\Omega_{\rm m}H(z)^2/H_0^2$ curve.}
}\label{fig:derived_tanh_wide}
\end{figure*}

In this section we present the constraints, at 68\% CL, for $h$ and $\Omega_{\rm m}$, along with a 
comparison of the best-fit of the model $-2\Delta \ln \mathcal{L_{\rm max}}$ and the Bayes Factor, with 
respect to $\Lambda$CDM, shown in~\cref{tabla_evidencias} for all the scenarios. 
Moreover, we show the posterior probability density functions, at 68\% and 95\% CL, for some quantities 
of interest in the interacting scenarios in~\cref{fig:derived_gp_nowide,fig:derived_gp_wide,fig:derived_tanh_nowide,fig:derived_tanh_wide}.

Beginning with the well known parameterizations $w$CDM and CPL, and by using all the combined 
datasets, i.e., BAO+H+SN, we obtained the following constraints on the parameters: 
$w_c=-0.99\pm0.06$, $w_0=-1.01\pm0.08$ and $w_a=0.12\pm0.47$. Their 
$-2\Delta\ln \mathcal{L_{\rm max}}$ are almost similar, among each other, with an improvement of 
$2.73$ for $w$CDM and $2.81$ for CPL with respect to the $\Lambda$CDM case, for one and two additional 
degrees of freedom, respectively (see also~\cref{tabla_evidencias}). 
The SSIK parameterization has, instead, three extra parameters with constraints $w_0=-0.91 \pm 0.05$, 
$\alpha = 0.97 \pm 0.79$ and $\sigma = 0.061 \pm 0.053$, and it presents a similar, albeit slightly 
better, fit of the data like the former parameterizations, with $-2\Delta\ln \mathcal{L_{\rm max}} = -3.12$. 
The evidences obtained favor $w$CDM over CPL and SSIK, with SSIK being the worst overall of the three 
parameterizations, which is not surprising as it has three extra parameters. Still when comparing any 
of the three scenarios with the standard cosmological model, even if they improve the fit of the data, 
the evidence is slightly against them, because models with additional parameters are more complex and 
therefore more penalized by the Occam's razor principle.  

Then we perform the reconstructions using five nodes interpolated via Gaussian Process for $\Pi(z)$ in two ways. 
One has an EoS parameter $w = -1$, for which we have $-2\Delta \ln \mathcal{L_{\rm max}} = -3.89$, 
and this represents an improvement of almost $2\sigma$ over the standard model. The other one with 
a variable EoS parameter $w_0 = -0.81 \pm 0.16$  with $-2\Delta \ln \mathcal{L_{\rm max}} = -4.22$, 
which is slightly better and suggests small deviations from $w=-1$. 
This stands out as the best model among the reconstructions. 
The main feature found in the functional posterior of $\Pi_{\rm DE}(z)$, shown in 
\cref{fig:derived_gp_nowide} and~\cref{fig:derived_gp_wide} (bottom right panel), is the presence 
of an oscillatory-like behavior around $\Pi_{\rm DE}=0$.
This behavior is present at the  1$\sigma$ level when $w=-1$ and becomes more pronounced when the DE EoS parameter is free to vary. In fact it is noticeable the presence 
of two maxima, one located at $z\sim 0.4$ and a more prominent one at $z\sim 2.3$, with deviations slightly 
outside the $1\sigma$ region. Additionally, there is also a local minimum at $z\sim 1.3$. Interestingly, 
all of them align closely with the positions of the BAO Galaxies and BAO Ly-$\alpha$ data, represented 
by the red error bars in the second panel of the figures.
The reconstruction of $\Pi(z)$ indicates (at $1\sigma$) more than one sign change in the flux of energy density 
transfer, that is, when the kernel switches from positive to negative the energy flow changes direction, 
i.e., in other words, at the beginning there is a flux of energy in the direction of  DE to DM, followed 
by a transition and thus the flux of energy reverses from DM to DE. 
The physical mechanism which makes this possible is beyond the scope of this work but it is important to note 
that similar results have been obtained before in~\cite{Cai:2009ht} with older versions of the data sets.

Once we have performed the reconstruction of $\Pi(z)$, we are able to extract some derived features,
shown in~\cref{fig:derived_gp_nowide,fig:derived_gp_wide}.
Here, we plot the functional posteriors for the quantities: $H(z)/(1+z)$ (corresponding to the expansion 
speed of the universe, i.e., $\dot a$ with $a$ being the scale factor), Solano’s dimensionless 
interaction function $I_Q(z)$, the deceleration parameter $q(z)$, the two effective EoS parameters 
($w_{\rm eff, \rm DM}$ and $w_{\rm eff, \rm DE}$), and both energy densities ($\rho_{\rm DM}/\rho_{\rm c,0}$ 
and $\rho_{\rm DE}/\rho_{\rm c,0}$). Both figures present a similar structure in the results, but the 
case where the EoS parameter is free to vary (Fig. \ref{fig:derived_gp_wide}) is a bit more pronounced, 
hence we focus on this case. 
The general form of $\Pi(z)$, including its oscillatory behavior,  is transferred to the derived 
functions. For instance, and as noted  before, the presence of maxima in the interaction kernel 
may be able to explain the BAO data. 
This can be seen in the panel with $H(z)/(1+z)$, which contributes to 
alleviate the BAO tension created between low redshift (galaxies) and high (Ly-$\alpha$) data, 
explored in~\cite{aubourg2015cosmological, Akarsu:2019hmw}. 
The fact that the general form of $H(z)/(1+z)$ changed, causes a displacement of its minimum 
value which in turn moves the beginning of the acceleration epoch to lower values of redshift, i.e. 
$q(z)=0$ at $z\sim 0.5$ at 68\% CL away from the $\Lambda$CDM value.
The main differences of the $\Lambda$CDM and the reconstructed IDE are accentuated  
on the functional posterior of the re-escalation function $I_Q(z)$. Here we notice the
existence of regions where the standard model remains outside the 68\% CL (1-$\sigma$),
which could motivate further studies of an interaction kernel with the presence of oscillations.
The general tendency of this function also resembles a previously obtained result
in~\cite{Bonilla:2021dql} with regards to the predominant negative values at late times. 
The last reconstructed derived features are the effective equations of state parameters 
and the energy densities. The effective EoS parameter of the DE, at low redshift, resembles a Quintom-like
behavior crossing the phantom-divide-line (PDL), viz., $w_\Lambda =-1$, multiple times; as studied in~\cite{Vazquez:2023kyx}.
A primary characteristic of the effective EoS parameter is  
exhibiting a pole (viz., ${\lim_{z\to z_\dagger^{\pm}}w_{\rm DE}(z)=\pm\infty}$ with $z_\dagger$ 
being the singular point). 
As studied in previous works~\cite{Akarsu:2020pka, Akarsu:2021fol, Escamilla:2021uoj, Ozulker:2022slu} 
this is necessary when a transition to a negative energy density is present, and this can 
also be easily verifiable by looking at the DE density, $\rho_{\rm DE}/\rho_{\rm c,0}$, which 
allows a transition to negative values at about $z\sim 2.3$. 
As a consequence of the interacting mechanism, the DM effective EoS parameter also shows some 
oscillations, although 
statistically in agreement with $w_{\rm DM} = 0$. 
Because the effective DM EoS parameter is a function of $Q(z)$ one finds that, if $Q(z) \neq 0$ 
then the DM would be no longer exhibit $\rho\propto a^{-3}$, i.e., no longer would behave like a fluid with 
a pressure identical to zero. This result is similar 
to the one obtained in~\cite{Kopp:2018zxp}, where the term used for a DM with a dynamic EoS parameter
was named Generalized dark matter (GDM). 
On the other hand, its energy density shows a tendency towards smaller values than $\Lambda$CDM (dashed line) at 
low redshifts, a possible transition to null or negative values at $z\sim 2.3$ and then larger 
values at higher redshifts.
This is a consequence of the changing direction in the energy transfer.
\\

\begin{table}[t!]
\caption{Mean values, and standard deviations,  for the parameters used throughout the reconstructions.
For each model, the last two columns present the Bayes Factor, and the 
$-2\Delta\ln \mathcal{L_{\rm max}} \equiv -2\ln( \mathcal{L_{\rm max}}_{,\Lambda \text{CDM}} / \mathcal{L_{\rm max,}}_i$) for fitness comparison. The datasets used are BAO+H+SN. 
Here $-2\ln \mathcal{L_{\rm max}}_{,\Lambda \text{CDM}} = 1429.7$, 
$\ln E_{\Lambda \text{CDM}}=-721.35 (0.14)$. 
}
\footnotesize
\scalebox{0.83}{%
\begin{tabular}{cccccc} 
\cline{1-6}\noalign{\smallskip}
\vspace{0.15cm}
Model &   EoS parameter  & $h$ &  $\Omega_{\rm m}$  &   $\ln B_{\Lambda \text{CDM},i}$  &  $-2\Delta\ln \mathcal{L_{\rm max}}$ \\
\hline
 \vspace{0.15cm}
$\Lambda$CDM & -1 &  0.683 (0.008) &  0.306 (0.013) &  0  &  0  \\
\vspace{0.15cm}
$w$CDM & $w_c$ &  0.675 (0.022) &  0.296 (0.016) &  1.51 (0.18)  &  -2.73 \\
\vspace{0.15cm}
CPL & $w_0+w_a(1-a)$ & 0.676 (0.023) & 0.298 (0.019) &  2.37 (0.19)  & -2.81  \\
\vspace{0.15cm}
SSIK  & $w_0$ & 0.681 (0.025) & 0.303 (0.027) &  3.82 (0.21)  & -3.12 \\
\hline
\hline
\vspace{0.15cm}
$\Pi_{\rm DE}$ GP & -1 &  0.684 (0.027) & 0.321 (0.032) &  8.61 (0.21)  & -3.89 \\
\vspace{0.15cm}
& $w_0$ &  0.687 (0.027) & 0.311 (0.024) &  8.01 (0.21)  & -4.22 \\
\hline
\vspace{0.15cm}
$\Pi_{\rm DE}$ bins & -1 & 0.684 (0.025) & 0.319 (0.027) &  5.69 (0.22) & -3.88 \\
\vspace{0.15cm}
& $w_0$ &  0.689 (0.027) & 0.314 (0.025) &  7.51 (0.22)  & -3.92 \\

\hline
\hline

\end{tabular}}

\label{tabla_evidencias}
\end{table}

\begin{table}[t!]
\caption{ 
Constraints at 68\% CL of the parameters for our model-independent reconstructions. The values for 
$\Pi_4$  are unconstrained for some of the cases, and for $\Pi_5$ for every case, which is expected 
given the lack of data in this redshift. 
}
\footnotesize
\scalebox{0.8}{%
\begin{tabular}{ccccccc} 
\cline{1-7}\noalign{\smallskip}
 \vspace{0.15cm}

Model &   $w_0$  & $\Pi_1$ &  $\Pi_2$  &  $\Pi_3$  &  $\Pi_4$  &  $\Pi_5$ \\
\hline
 \vspace{0.15cm}
$\Pi_{\rm DE}$ GP & $-1$           & $-0.01 (0.26)$ & $-0.26 (0.36)$ & $-0.29 (1.04)$ & $0.93 (5.34)$ &  unconstr. \\
\vspace{0.15cm}
  & $-0.81 (0.16)$ & $-0.78 (0.71)$ & $-0.16 (0.39)$ & $-0.22 (1.14)$ & $5.35 (5.82)$ & unconstr. \\

\hline
\vspace{0.15cm}
$\Pi_{\rm DE}$ bins & $-1$  & $0.04 (0.05)$ & $-0.61 (0.56)$ & $0.54 (3.16)$ & unconstr. & unconstr. \\
\vspace{0.15cm}
     & $-0.98 (0.09)$ & $0.02 (0.06)$ & $-0.43 (0.76)$ & $0.18 (3.55)$ & unconstr. & unconstr. \\

\hline
\hline

\end{tabular}}

\label{tabla_bestfit_ide}
\end{table}

\begin{figure*}
   \includegraphics[trim = 0mm  0mm 0mm 0mm, clip, width=18.cm, height=18.cm]{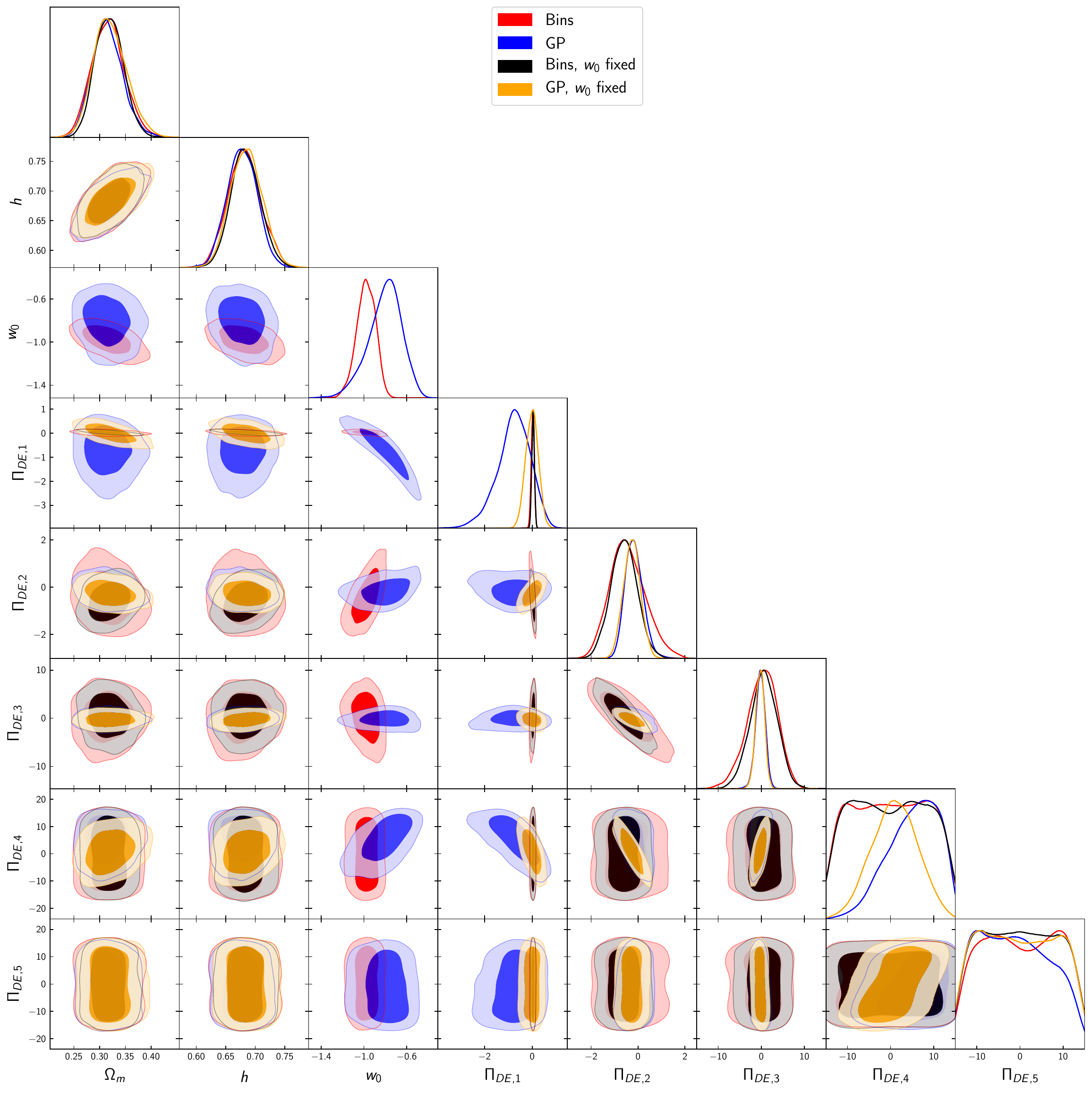}
   \caption{Triangle plot of every model-independent reconstruction. The parameter $w_{0}$ is only present in two of the reconstructions and it is correlated with $\Pi_{{\rm DE},1}$ when using a GP to perform the interpolation. 
   }\label{fig:triangleplot_all}
\end{figure*}

Next we present the results of the reconstruction using bins with~\cref{bin_equation} instead of GP. 
The functional posteriors can be seen in~\cref{fig:derived_tanh_nowide} ($w=-1$) and 
\cref{fig:derived_tanh_wide} ($w = w_0$). They look quite similar to the general features of the GP 
counterparts. When having $w = -1$ we obtain a $-2\Delta \ln \mathcal{L_{\rm max}} = -3.88$, and if the 
DE EoS parameter is allowed to vary  we get $w_0 = -0.98\pm 0.09$ and $-2\Delta\ln \mathcal{L_{\rm max}} = -3.92$
with respect to the $\Lambda$CDM scenario, improving the fit of the data and also performing 
slightly similar the reconstruction made with the GP interpolation. 
The results for this case also present oscillations around the null value of the interaction kernel 
$\Pi_{\rm DE}=0$ which, again, indicates more than one shift in the direction of energy density transfer. 
However, by using bins, the oscillations are noisier and thus more difficult to spot than in the one 
performed using GP; the results of the two cases, with fixed or varying EoS parameter, are very 
similar to each other. As far as the reconstructed derived features, we have a similar behavior to 
that found with GP. The re-escalation function $I_Q(z)$, for example presents some oscillatory-like 
behavior, that is less pronounced than the GP case, but lacks the first peak at
redshift $z \sim 0.5$. The Hubble parameter presents a horizontal flat region (darker green).
However, due to the larger confidence contours, it causes the existence of a region 
where the deceleration parameter equals zero, $z\sim 0.5-1.2$.
Finally, the effective DE EoS parameter presents again a pole, but in this case closer to $z=2$, {which indicates that the DE density, or $\rho_{\rm DE}(z)/\rho_{\rm c,0}$, is allowed to transit to negative values};
and the effective DM EoS parameter shows deviations from zero at more than 1$\sigma$ level.
These similar behaviors were expected as both model-independent reconstructions have similar 
degrees of freedom and the demeanor in which the nodes/bins are interpolated also have some 
visual similarities (as seen in~\cref{fig:comparison_tanh_gp} depending on the smoothness of the bins). {However, it is crucial to emphasize here that while their similarities are noteworthy, their differences are of equal importance. We will discuss this point in more detail at the end of this section.}

In~\cref{tabla_evidencias} we have the mean values and standard deviations for our parameter 
estimation procedure. Every model-independent reconstruction, regardless of its improvement in the fit of the data, 
presents a worse Bayes' Factor when compared to $\Lambda$CDM, because additional degrees of freedom 
are penalized by the Occam's razor principle. In~\cref{fig:triangleplot_all} we plot the 1D and 
2D marginalized posteriors of the parameters corresponding to $\Pi(z)$ and in~\cref{tabla_bestfit_ide} 
we report their 
constraints at 68\% CL, where the error is shown in parenthesis. The parameter $\Pi_1$, which is located in $z=0$ for GP, is clearly better 
constrained when taking $w=-1$, although its constraint is around $\Pi_1 = 0$, which indicates 
that, without a variable EoS parameter, it is pretty much forced to behave as $\Lambda$CDM at low redshifts.

When allowing variations on $w_0$, we note a separation from a $\Lambda$CDM-like behavior of around 
$1.5\sigma$ in $\Pi_1$ for GP. In contrast, when using bins this parameter is well constrained 
with or without a varying $w_0$. This happens because each bin spans a range ($\Delta z = 0.6$ 
in this case) and, specifically the first bin, is fitting all the available data in $0<z<0.6$ 
with a single step function making it very constrained, unlike its GP counterpart which uses both 
$\Pi_1$ and $\Pi_2$ (interpolated in $0<z<0.75$). Another interesting observation is that the 
restriction in $\Pi_1$ is reflected in the posterior of $w_0$, allowing it to be higher than $-1$ 
and presenting a negative correlation with $\Pi_1$ when using GP. The parameter $\Pi_3$ on the other hand, 
appears to be more constrained with GP than with bins. We can also see from the marginalized 1D 
posteriors that the parameter $\Pi_4$ is loosely constrained when using GP but unconstrained with 
bins. 
This different behavior could be attributed to the slight correlation imposed by the GP 
method, but also when $w_0$ 
is allowed to vary we see it is correlated with $\Pi_4$, and at the same time $\Pi_4$ is 
correlated with $\Pi_1$ which is also constrained.
$\Pi_5$ is completely unconstrained in all cases, but this was expected 
given the lack of data in this region ($z>2.4$). {Despite the significant findings presented above, it should be noted that the standard $\Lambda$CDM model still remains a viable option within the 2$\sigma$ confidence level, which means that we cannot definitely exclude it with the data we used in this study, and we need additional and more precise data sets to be able to say anything solid about this possibility.} {Let us continue with a brief discussion of the differences and similarities between the findings from the two different reconstruction approaches we used. For instance, it can be easily seen that certain characteristics are more evident in the GP reconstruction than in the binning method. This discrepancy could be attributed to the inherent correlations existing within GP among nodes, a correlation that is subtly reflected in the confidence contours shown in~\cref{fig:triangleplot_all}. These correlations seem to favor the GP approach, which is evidenced by a better fit of this approach to the data as can be seen in~\cref{tabla_evidencias}.
To reconcile these discrepancies among the approaches, a straightforward solution involves increasing the number of parameters, thereby achieving higher resolution. Nonetheless, this approach introduces the challenge of potential overfitting of specific characteristics and underfitting of others. 
To counterbalance this trade-off, we may need to incorporate a correlation function into the the binning method~\cite{zhao2017dynamical,Escamilla:2021uoj}, but the consideration of this is beyond the scope of the present work although it might be a promising direction for future investigations. It seems reasonable to conclude from this discussion that some of the observed features may be influenced by the chosen reconstruction method, but certain general characteristics persist regardless of the approach. 
These enduring traits include the oscillatory behavior at $1\sigma$, the asymptotic behavior of the effective EoS and the possibility of a transition to a negative DE density.}

{We conclude this section by commenting on one of the most interesting findings of our study, the possibility of the existence of a DE that can take negative density values at high redshifts (viz., for $z\gtrsim2$), regardless of the approach used. Although this possibility may seem physically unexpected and challenging, it is not a new finding in our study and has been studied in the previous literature, especially recently, to address the cosmological tensions such as the $H_0$ and $S_8$ tensions; see, for instance, Refs.~\cite{Akarsu:2019hmw, Akarsu:2021fol,Akarsu:2022typ,Akarsu:2023mfb,Adil:2023exv} considering models that suggest such a transition at $z\sim2$ from their observational analysis and references therein for further reading. This type of DE behavior was also predicted in a model-independent manner in a recent study~\cite{Escamilla:2021uoj} that directly reconstructed the DE density. Our findings here present a noteworthy distinction with this recent study, as in the current study we achieved a similar behavior by incorporating an interacting dark sector (dark matter+dark energy) instead of employing a direct reconstruction of the DE interacting only gravitationally. This observation holds significance as it indicates that the data sets consistently favor (or at the very least allow for) a negative DE density for $z\gtrsim2$, irrespective of the method employed. This finding, combined with the model's potential to address certain cosmological tensions (as extensively discussed in~\cite{Akarsu:2022typ, Akarsu:2023mfb}), emphasizes the notion that this model emerges as a promising alternative to the standard $\Lambda$CDM model.}

\section{Conclusions}\label{section:conclusions}

Throughout this paper,  we performed model-independent reconstructions of the interaction kernel between 
DM and DE by implementing an interpolation with both Gaussian Process and bins joined via hyperbolic tangents, 
using the SimpleMC code along with the Nested Sampling algorithm. 
The main results showed that particular features, such as oscillations are present, but they remain still 
statistically consistent with the $\Lambda$CDM model. 
By using these reconstructions some derived functional posteriors were also obtained, which inherit 
the general characteristics of $\Pi(z)$. 
These oscillatory features can be more clearly observed through the reescalation function introduced in~\cite{Solano:2012zw}, 
and it is worth noting that similar shapes were also found in a model-independent reconstruction in~\cite{Cai:2009ht};
(see also \cite{Akarsu:2022lhx}, suggesting that, in the relativistic cosmological models that 
deviate from $\Lambda$CDM, dark energies are expected to exhibit such behaviors for the consistency with CMB data).
We noticed the Hubble parameter was slightly modified in order to alleviate the tension created between 
low and high redshift BAO data (reflected in the improvement of the fit) which also causes a shift, to later times, 
for the beginning of the acceleration epoch.
When plotting the functional posterior of the DE effective EoS parameter we observed a quintom-like behavior
at low redshift, with a preference zone of the $68\%$ confidence contour away from the $\Lambda$CDM.
Additionally, we observe the presence of a pole at about $z\sim 2.3$ recovering a shape with an asymptote, 
proposed and studied  in other works~\cite{Akarsu:2020pka, Akarsu:2021fol, Escamilla:2021uoj,
gomez2015background, wang2019searching}. 
This particular shape is required when having a DE energy density that presents a transition from positive to negative energy density or vice versa.  This transition is shown to be possible in the 68\% contour 
of the derived DE energy density. 
Last but not least, an important implication of these reconstructions is seen with~\cref{eq:effective_eos}. 
We found a non negligible interaction kernel, thus the effective behavior of DM, at the largest scales,  
may not be described by a perfect pressure-less fluid but something around it.

Despite the positive outcomes observed in the fit of the data, we cannot ignore the Bayes' Factors. 
As our model-independent method introduces several new parameters, it is expected to be in 
disadvantage when compared to the concordance model. 
To achieve improved results without disregarding our findings, it would be advisable to consider a 
new parameterization or a change in the basis with a considerably reduced number of parameters, that can 
take into account the new found features.

\begin{acknowledgments}

\"{O}.A. acknowledges the support by the Turkish Academy of Sciences in the scheme of the Outstanding Young Scientist Award  (T\"{U}BA-GEB\.{I}P). \"{O}.A. is supported in part by TUBITAK grant 122F124. E.D.V. is supported by a Royal Society Dorothy Hodgkin Research Fellowship. L.A.E. was supported by CONACyT M\'exico. J.A.V. acknowledges the support provided by FOSEC SEP-CONACYT Investigaci\'on B\'asica A1-S-21925, FORDECYT-PRONACES-CONACYT/304001/2020 and UNAM-DGAPA-PAPIIT IN117723. This article is based upon work from COST Action CA21136 Addressing observational tensions in cosmology with systematics and fundamental physics (CosmoVerse) supported by COST (European Cooperation in Science and Technology).
\end{acknowledgments}


\bibliographystyle{apsrev4-1}
\bibliography{bibliography.bib}


\end{document}